\documentclass[10pt,aps,prr,twocolumn,superscriptaddress,floatfix]{revtex4-2}

\usepackage{graphicx,amsmath,amsfonts,amssymb,mathtools,bm,times}
\usepackage{relsize,epstopdf,upgreek,color,xcolor,rotating,tabularx,booktabs,setspace,makecell,cancel}
\usepackage{algorithm,algpseudocode,appendix}

\usepackage[colorlinks=true,linkcolor=blue,citecolor=blue,urlcolor=blue]{hyperref}

\newcommand{\rhoq}{\rho_\textsc{q}}
\newcommand{\rhob}{\rho_\textsc{b}}
\newcommand{\argp}[1]{\left(#1\right)}
\newcommand{\args}[1]{\left[#1\right]}
\newcommand{\argc}[1]{\left\{#1\right\}}
\newcommand{\ketbra}[2]{\ket{#1}\!\!\bra{#2}}

\newcommand{\ket}[1]{\left| #1\right>}
\newcommand{\bra}[1]{\left< #1\right|}
\newcommand{\inner}[2]{\langle #1|#2\rangle}
\newcommand{\opinner}[3]{\langle #1|#2|#3\rangle}

\newcommand{\rvec}[1]{\pmb{#1}}

\newcommand{\tr}[1]{\mathrm{tr}\!\left\{#1\right\}}

\newcommand{\Tr}[1]{\mathrm{Tr}\{#1\}}
\newcommand{\D}{\mathrm{d}}
\newcommand{\I}{\mathrm{i}}

\newcommand{\E}[1]{\mathrm{e}^{\mbox{\footnotesize$#1$}}}

\newcommand{\MEAN}[1]{\left<{#1}\right>}
\newcommand{\AdA}{a^\dagger a}

\newcommand{\AVG}[2]{\underset{#1}{\displaystyle{\mathop{\mathbb{E}}}}\!\left[#2\right]}

\newcommand{\appropto}{\mathrel{\vcenter{
			\offinterlineskip\halign{\hfil$##$\cr
				\propto\cr\noalign{\kern2pt}\sim\cr\noalign{\kern-2pt}}}}}

\newcommand{\etaT}{\eta}
\newcommand{\etal}{\eta}
\newcommand{\etad}{\eta^\prime}
\newcommand{\Us}{U_\mathrm{s}}

\newcommand{\khat}{\widehat{\mathbf{k}}}
\newcommand{\nhat}{\widehat{\mathbf{n}}}
\newcommand{\SL}{C_\textsc{l}}

\begin{document}
	
	\author{Saurabh U. Shringarpure}
	\affiliation{NextQuantum Innovation Research Center, Department of Physics and Astronomy,  Seoul National University, Seoul 08826, South Korea}
	
	\author{Siheon Park}
	\affiliation{NextQuantum Innovation Research Center, Department of Physics and Astronomy,  Seoul National University, Seoul 08826, South Korea}
	
	\author{Sungjoo Cho}
	\affiliation{NextQuantum Innovation Research Center, Department of Physics and Astronomy,  Seoul National University, Seoul 08826, South Korea}
	
	\author{Yong Siah~Teo}
	\email{ys\_teo@snu.ac.kr}
	\affiliation{NextQuantum Innovation Research Center, Department of Physics and Astronomy,  Seoul National University, Seoul 08826, South Korea}
	
	\author{Hyukjoon Kwon}
	\email{hjkwon@kias.re.kr}
	\affiliation{School of Computational Sciences, Korea Institute for Advanced Study, Seoul 02455, South Korea}
	
	\author{Srikrishna Omkar}
	\email{omkar.shrm@gmail.com}
	\affiliation{NextQuantum Innovation Research Center, Department of Physics and Astronomy,  Seoul National University, Seoul 08826, South Korea}
	
	\author{Hyunseok Jeong}
	\email{h.jeong37@gmail.com}
	\affiliation{NextQuantum Innovation Research Center, Department of Physics and Astronomy,  Seoul National University, Seoul 08826, South Korea}
	
	\title{Ballistic bosonic noise suppression with hybrid qumode-qubit rotation gates}
	
	\begin{abstract}
		
		\mbox{Noise suppression is of paramount importance for reliable quantum information processing and computation.} 
		We show that for any single-mode bosonic code (qumode) corrupted by thermal~noise at rate~$\eta$ and mean \mbox{excitation}~$\bar{n}$, a hybrid continuous-discrete-variable~(CV-DV) interferometer using only a single qubit ancilla~(DV) and two controlled~Fourier~(CF) gates sandwiching the noise channel suppresses its effects to $\mathcal{O}(\eta^2)$ \emph{without} any active error correction or destructive measurements of the encoded state and with high success probabilities~$>0.5$ if~$\eta(1+\bar{n})<0.5$. This suppression scheme works by conditionally monitoring the photon-number parities after the interferometer. Bosonic codes with two logical states of the same photon-number parity (like-parity codes) are \emph{completely resilient} to DV amplitude- and phase-damping ancilla noise. 
		For such codes, the interferometer simplifies to the use of a qumode rotation gate and a \emph{single} CF~gate. This presents a clear advantage of our CF-gate-based error suppression scheme over previously-proposed ``bypass'' protocols, where qubit information transferred to the DV mode is readily corrupted by damping~noise. Finally, we present a simple extension to direct communication of qumode states between two parties over a noisy channel using a preshared DV entangled state, by implementing a CF gate in the first laboratory and its inverse in the other. Such a communication protocol achieves a similar fidelity performance at the same success rate as the single-party case, but with greater resilience to the ancilla noise than DV~teleportation.
		Resource-efficient multi-qubit codes that depend on a few essential long-range interactions can benefit from it. 
	\end{abstract}
	
	\maketitle
	
	\emph{Introduction.---}Strong, controlled interactions between two quantum systems are essential components of quantum information processing tasks. Rabi model~\cite{Rabi1936:On, Rabi1937:Space} and its variations in different regimes~\cite{Jaynes1963:Comparison, Brune1996:Observing, Schleich2001:Quantum, Zueco2009:Qubit-oscillator, Vaaranta2025:Analytical} describe atom-light-like couplings~\cite{Frisk2019:Ultrastrong, Heeres2015:Cavity, Krastanov2015:Universal}, where one is characterized by discrete variables~(DV), and the other by continuous variables~(CV). These have been employed in various quantum computing platforms such as cavity quantum electrodynamics~\cite{Haroche1989:Cavity, Raimond2001:Manipulating}, circuit quantum electrodynamics~\cite{Wallraff2004:Strong, Blais2004:Cavity, Blais2021:Circuit}, and trapped ions~\cite{Leibfried2003:Quantum, Lv2018:Quantum}.
	Linear optics with measurement back action provides an alternative route to realizing nonlinear interactions~\cite {Knill2001:Scheme}.
	
	While bosonic-encoded quantum information (qumodes) as standing waves inside superconducting microwave cavities or waveguides have been explored~\cite{Leghtas2013:Hardware-Efficient, Leghtas2013:Deterministic, Xiang2013:Hybrid, Ofek2016:Extending, Sivak2023:Real, Ni2023:Beating, Campagne2020:Quantum, Eickbusch2022:Fast, Gao2021:Practical}, there is a case for flying qumodes that enable distributed computing~\cite{Duan2004:Scalable, Duan2005:Robust, Spiller2006:Quantum, Kimble2008:Quantum, Kimble2008:Quantum, Reiserer2014:Quantum, Tiecke2014:Nanophotonic, Reiserer2015:Cavity, Reiserer2015:Controlled, Hacker2019:Deterministic, Bravyi2024:High, Yang2025:Deterministic}. They can also encode error-correctable quantum information, in a hardware-efficient alternative to DV-only quantum computing~\cite{ Albert2018:Performance, Leghtas2013:Hardware-Efficient}. Moreover, hybrid CV-DV systems that use the benefits of both kinds of systems have various applications~\cite{Park2012:Quantum, Lee2013:Near-deterministic, Kwon2013:Violation, Jeong2014:Generation, Generation2015:Kwon, Bang2015:Quantum, Jeong2016:Quantum, Omkar2020:Resource-Efficient, Omkar2021:Highly, Gan2020:Hybrid, Lee2024:Fault-tolerant, Bose2024:Long-distance, Jeon2025:Amplifying, Li2023:Memoryless, Li2024:Performance, Liu2025:Hybrid, Bera2025:Sharing}. 
	However, noise hinders all such desired applications. 
	Therefore, it is essential to suppress noise as much as possible at the physical level, for example, to reduce resource overheads of error correction~\cite{Teo2025:Linear, Omkar2020:Resource-Efficient, Lee2024:Fault-tolerant}, despite its performance breaking even with the uncorrected case~\cite{Sivak2023:Real, Ni2023:Beating}. 
	
	Losses pose the greatest challenge to qumodes, along with thermal excitation, random displacement, and~dephasing~\cite{Teo2025:Linear}, and several methods have been proposed to manage CV noise. The ``bypass'' protocol~\cite{Park2022:Slowing} was shown to slow decoherence deterministically by transferring quantum information from the qumode to the DV ancilla before it encountered the CV noise. Linear optics is effective at canceling the effects of photon losses~\cite{Taylor2024:Quantum}, thermal excitation, and random displacement noise~\cite{Teo2025:Linear} through error mitigation but at exorbitant sampling costs~\cite{Quek2024:Exponentially}. 
	While linear optics can suppress bosonic dephasing~\cite{Marshman2018:Passive, Swain2024:Improving, Teo2025:Linear}, active squeezing can address losses and thermal noise~\cite{LeJeannic2018:Slowing, Brewster2018:Reduced, Schlegel2022:Quantum, Provaznik2025:Adapting,Pan2023:Protecting, Rousseau2025:Enhancing} but relies on the asymmetrical phase space properties. Beyond photon losses~\cite{Ralph2011:Quantum, Micuda2012:Noisless}, direct suppression of thermal noise with only linear optics is elusive~\cite{Teo2025:Linear}. 
	
	In this Letter, we propose an experimentally feasible scheme, with high success rates, to suppress bosonic photon losses and thermal noise corrupting a qumode. It features an interferometer that entangles the qumode with a single DV qubit ancilla using only two hybrid qumode-qubit rotation gates, realizable with dispersive Rabi interactions~\cite{Schleich2001:Quantum, vanLoock2006:Hybrid, Reiserer2014:Quantum, Hacker2019:Deterministic}. We show that when the rotation gates are controlled-Fourier (CF) gates, not only will first-order noise effects be fully suppressed, but the error-suppression performance of our CF-based protocol is \emph{completely resilient} to DV phase- and amplitude-damping noise when bosonic codes of identical photon-number-parity codewords are~considered. Such resilience originates from the unique parity-monitoring mechanism underlying our CF-based protocol without directly transferring quantum information into the DV~ancilla, which is in stark contrast to earlier methods~\cite{Park2022:Slowing, Hastrup2022:Universal, Park2023:Quantum, Park2025:Quantum}. Our proposed scheme is therefore both \emph{ballistic} (no adaptive feedback required) and the required gate settings are \emph{code-agnostic}. Significant error suppression (near-unit suppression fidelity) is observed for typical bosonic~codes. Extension to the Gaussian displacement noise is~straightforward. We also apply the CF-based suppression scheme in quantum communication~\cite{Li2023:Memoryless, Li2024:Performance}, where we prove that noise corrupting a two-party qumode-sharing channel can be significantly suppressed using remotely entangled DV qubits as a~resource.
	
	\begin{figure}[t]
		\centering
		\includegraphics[width=1\columnwidth]{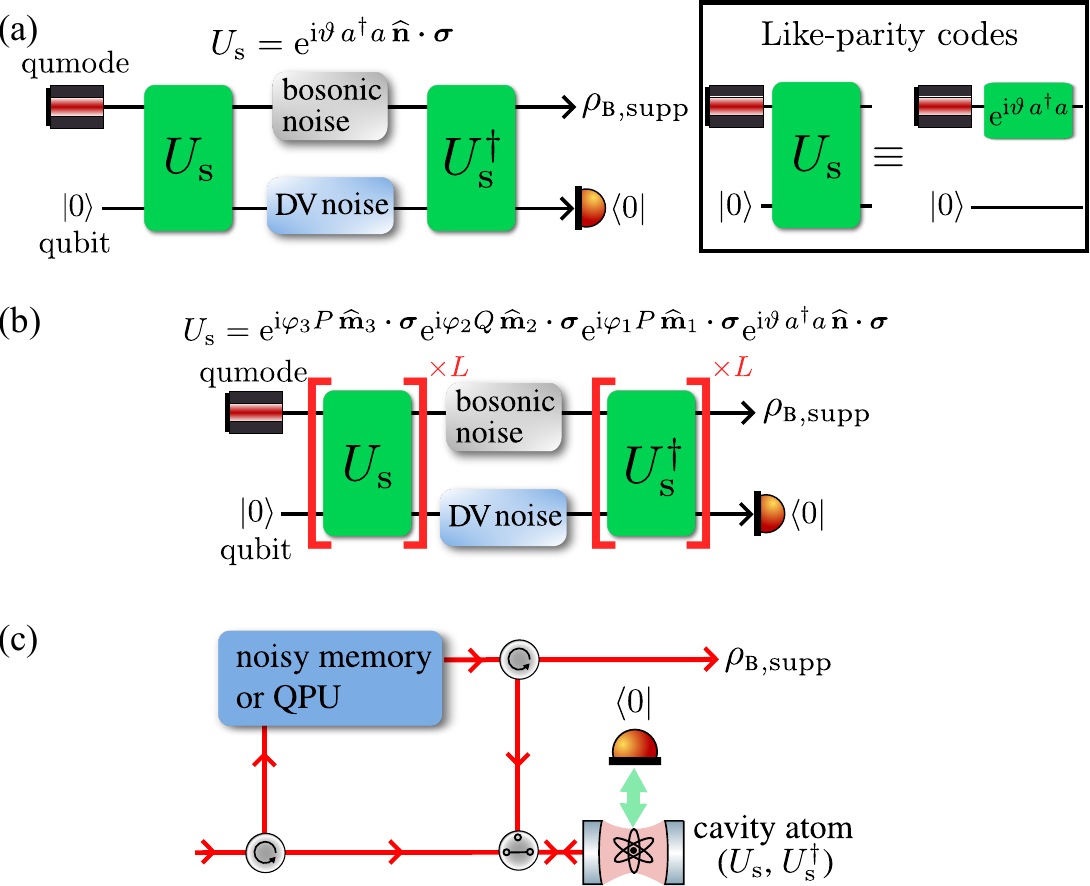}
		\caption{\label{fig:protocols} 
			Bosonic noise suppression with (a) conditional rotation~(condrot) gates and (b) a series of compositions of conditional displacements and the conditional rotations gates. (c) An optical circuit implementing conditional rotations using the dispersive interactions between a single atom in a cavity and traveling waves~\cite{Reiserer2015:Controlled} with two circulators and a time-dependent optical switch to redirect the beams onto a cavity containing an atom to implement the noise suppression unitaries $U_\mathrm{s}$ and $U_\mathrm{s}^\dagger$. The atomic state is read out~(green~arrow)~\cite{Hacker2019:Deterministic,Nunn2021:Heralding} and the CV output $\rho_{\textsc{b},\text{succ}}$ is heralded on measuring the ground state. For bosonic codes with logical states of identical parity, the first hybrid gate can be replaced with a local qumode rotation as shown by the inset in~(a).
		}
	\end{figure}
	
	\emph{Noise sources.---}We model CV noise of rate $\eta$ as thermal noise with mean photon number $\bar{n}$ [including photon loss ($\bar{n}=0$)], which is viewed as a quantum-limited amplification channel of gain $G=1+\eta \bar{n}$ following a pure loss one of rate \mbox{$\mu=1-(1-\eta)/G$}. As a side remark, we point out that Gaussian displacement channels can be similarly decomposed into pure loss and a quantum-limited amplifier~\cite{Noh2019:Quantum}. As such, all results in this Letter apply also to this kind of~noise. The Kraus representations of pure loss and the amplifier are~\cite{Ivan2011:Operator-sum, Gagatsos2017:Bounding, Noh2019:Quantum}, \mbox{$
			\mathcal{N}_\mathrm{loss}[\rho_\textsc{b}]=\,\sum_{l\geq0}\frac{\mu^l}{l!}(1-\mu)^{\frac{a^\dag a}{2}}\,a^l\,\rho_\textsc{b}\,a^{\dag\,l}\,(1-\mu)^{\frac{a^\dag a}{2}}
			$} 
		and 
		\mbox{$  
			\mathcal{N}_\mathrm{amp}[\rho_\textsc{b}]	=\sum_{k\geq0}\frac{(1-G^{-1})^k}{k!\,G}\,a^{\dag k}\,G^{-\frac{a^\dag a}{2}}\,\rho_\textsc{b}\,G^{-\frac{a^\dag a}{2}}\,a^{k},
			$}
		\hfill\null
		respectively, were $a$ and $a^\dagger$ are the bosonic ladder operators and $\rho_\textsc{b}$ is the density operator of the bosonic~mode.
		
		The DV phase and amplitude damping channels on the ancilla qubit state~$\rho_\textsc{q}$ have the form $\mathcal{N}_{\mathrm{damp}}[\rho_\textsc{q}]=\sum_{j=0}^1K_j\rho_\textsc{q}\,K_j^\dagger$, where \mbox{$K_0=\ket{0}\!\!\bra{0}+\sqrt{1-p}\ket{1}\!\!\bra{1}$} and \mbox{$K_{1}=\sqrt{p}\ket{1}\!\!\bra{1}$} and \mbox{$\sqrt{p}\ket{0}\!\!\bra{1}$} respectively~\cite{Nielsen2010:Quantum}. We shall consider amplitude and phase damping noise of equal strengths~$p$ on the ancilla throughout this~Letter. 
		
		\emph{Single ancilla-assisted suppression.---}The hybrid interferometer we consider here applied to the qumode is represented by the output state $\rho_{\textsc{b},\mathrm{supp}}\propto\, \bra{0}\Us^\dag\,\mathcal{N}[\Us\,\rho_\textsc{b}\otimes\ket{0}\!\!\bra{0}\Us^\dag]\,\Us\ket{0}$, where $U_\mathrm{s}$ is the suppression unitary, $\mathcal{N}$ is the noise affecting the hybrid CV-DV state, and the normalization gives the success~probability of the protocol. The ancilla is initialized and finally, conditionally measured in the same state~$\ket{0}\!\!\bra{0}=(1+\khat\bm{\cdot}\bm{\sigma})/2$, unless stated otherwise. For a two-level system like an atom, this is the ground state and a natural choice for the ancilla qubit, as it remains stable \mbox{under~damping~noise}. 
		The suppression unitary we consider is $\Us=\E{\I\vartheta\,a^\dag a \,\widehat{\mathbf{n}}\cdot\rvec{\sigma}}$, where $\rvec{\sigma}$ is the column of Pauli operators. This is equivalent to two additional (DV) single-qubit Hadamard gates sandwiching a conditional $Z$-rotation~\cite{Nielsen2010:Quantum}.
		
		For any $\rho_\textsc{b}$ and small $\etaT$, the configuration corresponding to the CF gate,~$\vartheta=\pi/2$ and $\nhat\perp\khat$, completely cancels the effects due to all the unwanted pairs of photon loss and gain events that alter the photon-number parity, assuming perfect DV~ancilla. Such cancellation of paired losses and gains leads to noise suppression for \emph{any} in put state~$\rho_\mathrm{in}$, manifesting as the absence of $\mathcal{O}(\eta)$ terms in the fidelity of the output $\rho_\mathrm{out}$. By averaging over Haar-random~\cite{Harrow2013:Church, Mele2024:Introduction} pure qubit input states encoded on the \mbox{orthonormal} bosonic logical kets as $\ket{\psi_\mathrm{in}}=c_0\ket{0_\textsc{l}}+c_1\ket{1_\textsc{l}}$, the \emph{average fidelity} $\overline{\mathcal{F}}=\AVG{c_0,c_1}{\opinner{\psi_{\mathrm{in}}}{\rho_\mathrm{out}}{\psi_{\mathrm{in}}}}$, over real~$c_0$ and complex~$c_1$ on the surface of the qubit Bloch~sphere, 
		reflects the average quality of encoded output physical states, improving which also improves that of logical ones upon further error~correction. 

        \begin{figure}[t]
			\centering
			\includegraphics[width=\columnwidth]{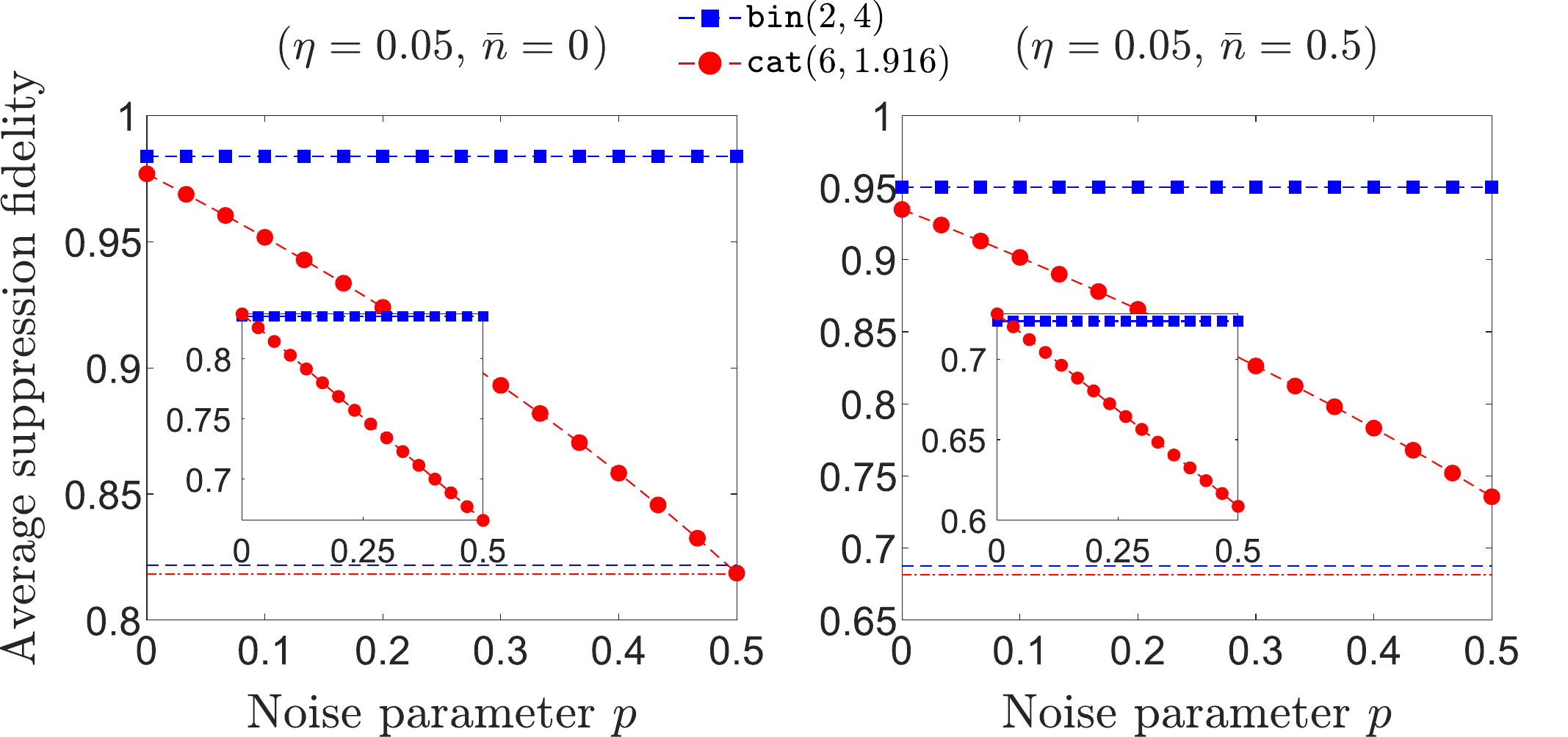}
			\caption{\label{fig:likevsopp}
				Comparison of average suppression performance with conditional Fourier~(CF) gates between a like (even)-parity code $\texttt{bin}(2,4)$ and an opposite-parity bosonic code: $\texttt{cat}(6,1.916)$, both of which have similar average Gaussian moments of the photon-number distribution~$(\langle n\rangle\cong 4$, \mbox{$\langle n^2 \rangle\cong 20$}, \mbox{$\langle a^2\rangle=0)$} with respect to the state~$C_\textsc{l}$. The former is more resilient to the composite amplitude and phase damping qubit noise of equal strengths~$p$. The dashed and dot-dashed lines represent their respective, roughly \mbox{identical~performances}. Insets show average success~probability. }
		\end{figure}
		
		For our CF-based suppression protocol, the \emph{average suppression fidelity} (Sec.~{A} of Supplemental Material (SM)~\cite{SM}) 
		\begin{align}
			&\,\overline{\mathcal{F}}_\mathrm{supp}\cong1-\etaT^2\bigg\{\bar{n}^2+3\,\left(\bar{n}+\frac{1}{2}\right)^2\,\tr{\SL\,(a^\dag a)^2}\nonumber\\
			&\,+\left(\bar{n}^2-\bar{n}-\frac{1}{2}\right)\tr{\SL\,a^\dag a}-\left[\frac{1}{6}+\frac{4}{3}(\bar{n}^2+\bar{n})\right]g(a^\dag a)\nonumber\\
			&\,-\left[\frac{1}{3}+\frac{2}{3}\left(\bar{n}^2+\bar{n}\right)\right]g(a^2)\bigg\}
			\label{eq:Fsupp}
		\end{align}
		is achieved for thermal noise, where $g(Y)=\tr{\SL\,Y\SL\,Y^\dag}+|\tr{\SL\,Y}\!|^2$ and
		$\SL=\big(\ket{0_\textsc{l}}\!\!\bra{0_\textsc{l}}+\ket{1_\textsc{l}}\!\!\bra{1_\textsc{l}}\big)/2$ is the normalized codespace identity. Note that the suppression protocol results in a ballistic error-rate scaling ($\eta^2$) of the infidelity~$1-\overline{\mathcal{F}}_\mathrm{supp}$. This offers an advantage over the unsuppressed case where the corresponding infidelity scales linearly with $\eta$~(Sec.~{C} of SM~\cite{SM}). The corresponding \emph{average success probability} reads (Sec.~{B} of SM~\cite{SM})
		\begin{equation}
			\overline{p}_\mathrm{succ}=\frac{1}{2}+\frac{1}{2(2G-1)}\tr{\SL
				\left(\frac{1-2\mu G}{2G-1}\right)^{\AdA}}.
			\label{eq:psucc}
		\end{equation}
		As $\SL\geq0$ and $G>1$, $\overline{p}_\mathrm{succ}>0.5$ when $\mu G<0.5$, which amounts to the condition $\eta(1+\bar{n})<0.5$ for thermal~noise.
		
		In what follows, we shall present concrete examples from the families of common single-mode bosonic codes~\cite{Albert2018:Performance, Grimsmo2020:Quantum, Teo2025:Linear}: (i) the cat codes $\texttt{cat}(n,\alpha)$ which are superpositions of $n$ coherent states of amplitude $\alpha$ on a ring~\cite{Cochrane1999:Macroscopically, Jeong2002:Efficient,Ralph2003:Quantum,Leghtas2013:Hardware-Efficient}, (ii) the binomial codes $\texttt{bin}(n,\kappa)$  which are superpositions of maximum $\kappa$, $n$-gapped Fock states distributed binomially~\cite{Michael2016:New} and (iii) the finite-energy approximate Gottesman--Kitaev--Preskill codes $\texttt{gkp}(\Delta)$ which are superpositions of displaced-squeezed states with a damping factor~$\Delta$~\cite{Gottesman2001:Encoding}. Our scheme, clearly, also protects the bosonic mode of hybrid-entangled states (such as $\ket{\alpha}\!\!\ket{0}+\ket{\beta}\!\!\ket{1}$) with one extra DV ancilla owing to linearity in the suppression~action~(see Sec. F of SM~\cite{SM}). 

        \begin{figure}[t]
			\centering
			\includegraphics[width=\columnwidth]{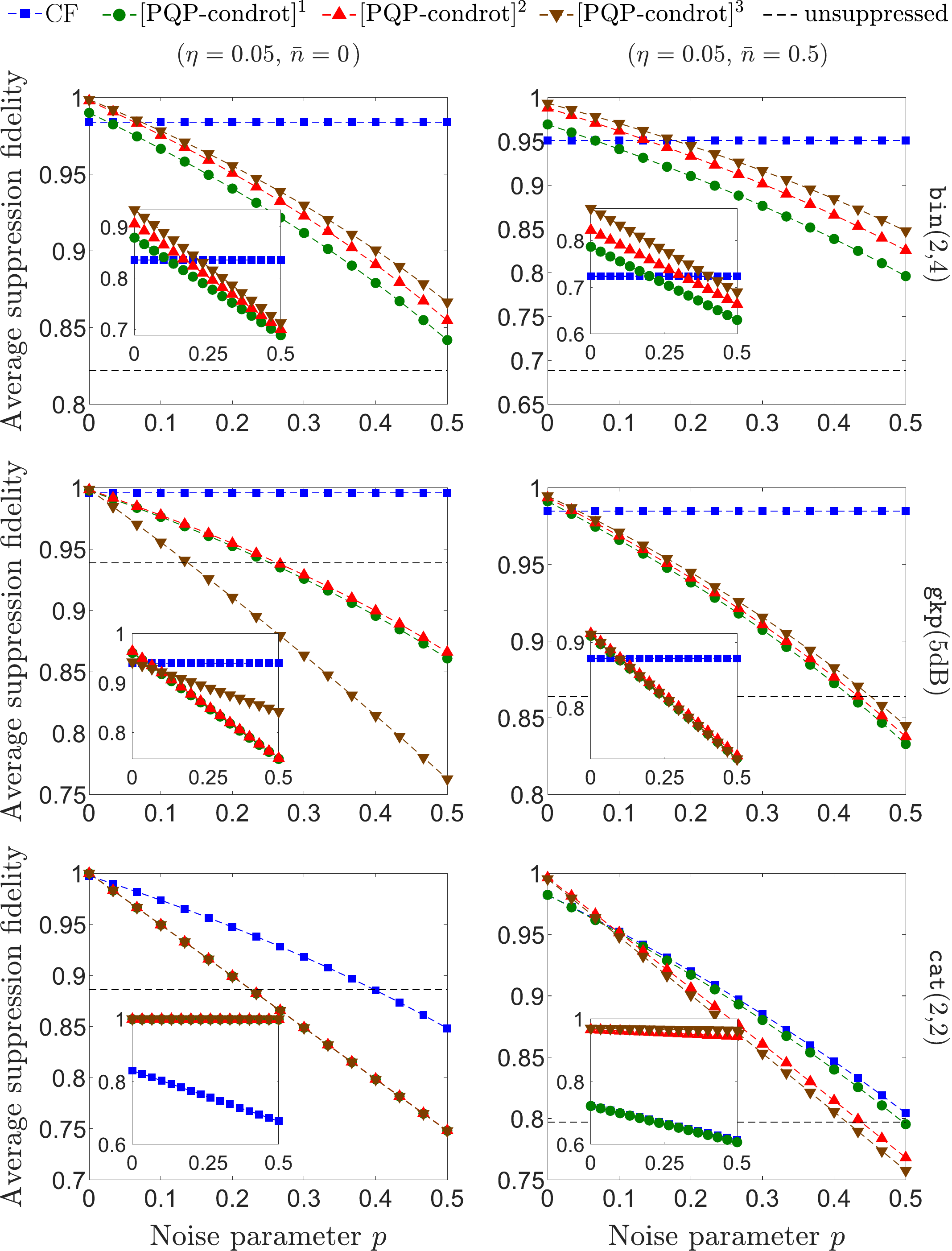}
			\caption{\label{fig:PQPcondrot_damp}  
				The CF-gate-only interferometer requires no information about the noise parameters. Its average suppression performance remains impervious to uncalibrated ancilla damping, contrary to a series of conditional displacement gates and conditional rotation gates~([PQP-condrot]$^L$) [see Fig.~\ref{fig:protocols}~(b)], numerically optimized for known noise parameters for photon loss~($\eta=0.05$) and thermal noise~($\eta=0.05$, $\bar{n}=0.5$). Insets show average success~probability.}
		\end{figure}  
        
		\emph{Complete resilience to damping.---}Consider families of bosonic codes stabilized by the parity operator, 
		\mbox{$[\ket{0_\textsc{l}}\!\!\bra{0_\textsc{l}},(-1)^{a^\dag a}]=0=[\ket{1_\textsc{l}}\!\!\bra{1_\textsc{l}},(-1)^{a^\dag a}]$}, 
		and that the logical states have either like or opposite photon-number parities. Since the first CF~unitary gate is given by $ U_\mathrm{s}=\cos(\pi\AdA/2)+\I\sin(\pi\AdA/2)\nhat\bm{\cdot}\bm{\sigma},$
		it is clear that only the first term survives for like (even)-parity codes as $\cos(\pi\AdA/2)=\sum_{k\geq0}(-1)^k\ket{2k}\!\!\bra{2k}$ lives in the even subspace. This implies that the ancilla remains in the ground state, unaffected. It remains stable as such under damping noise up to the action of the second unitary $U_\mathrm{s}^\dag$, 
		which subsequently flips the DV state \emph{only} if the bosonic noise alters photon-number~parity.
		
		As the second term plays no role, we can further simplify the setup by substituting the first conditional rotation with a single, local, bosonic rotation by $\pi/2$ for an identical performance. A similar argument applies to like (odd)-parity codes when the DV ancilla is initialized in the state $\nhat\bm{\cdot}\bm{\sigma}\ket{0}$ instead. In this case, the ancilla state after the first unitary is~$\ket{0}$ again, leading to the same robustness under DV damping~noise. 
		
		\emph{Distinction from the ``bypass" protocol.---}Numerical analysis found that additional \emph{conditional displacement gates} interleaved with conditional rotations can enhance the fidelity further if the DV~ancilla is nearly perfect, with some resilience to small DV noise, as illustrated by Fig.~\ref{fig:PQPcondrot_damp}.
		Such multiple conditional displacements have been employed to ``bypass" the CV noise by transferring the quantum information over to the DV system before noise~\cite{Park2022:Slowing, Hastrup2022:Universal}. However, ``bypass'' protocols, designed primarily for two- and four-component cat codes, are susceptible to DV damping noise and require additional gate resources. Moreover, by transferring CV quantum information into DV systems, which are typically stationary such as atoms in a cavity, the unique practical advantages of traveling waves are forfeited. 
		
\begin{figure}[t!]
			\centering
			\includegraphics[width=\columnwidth]{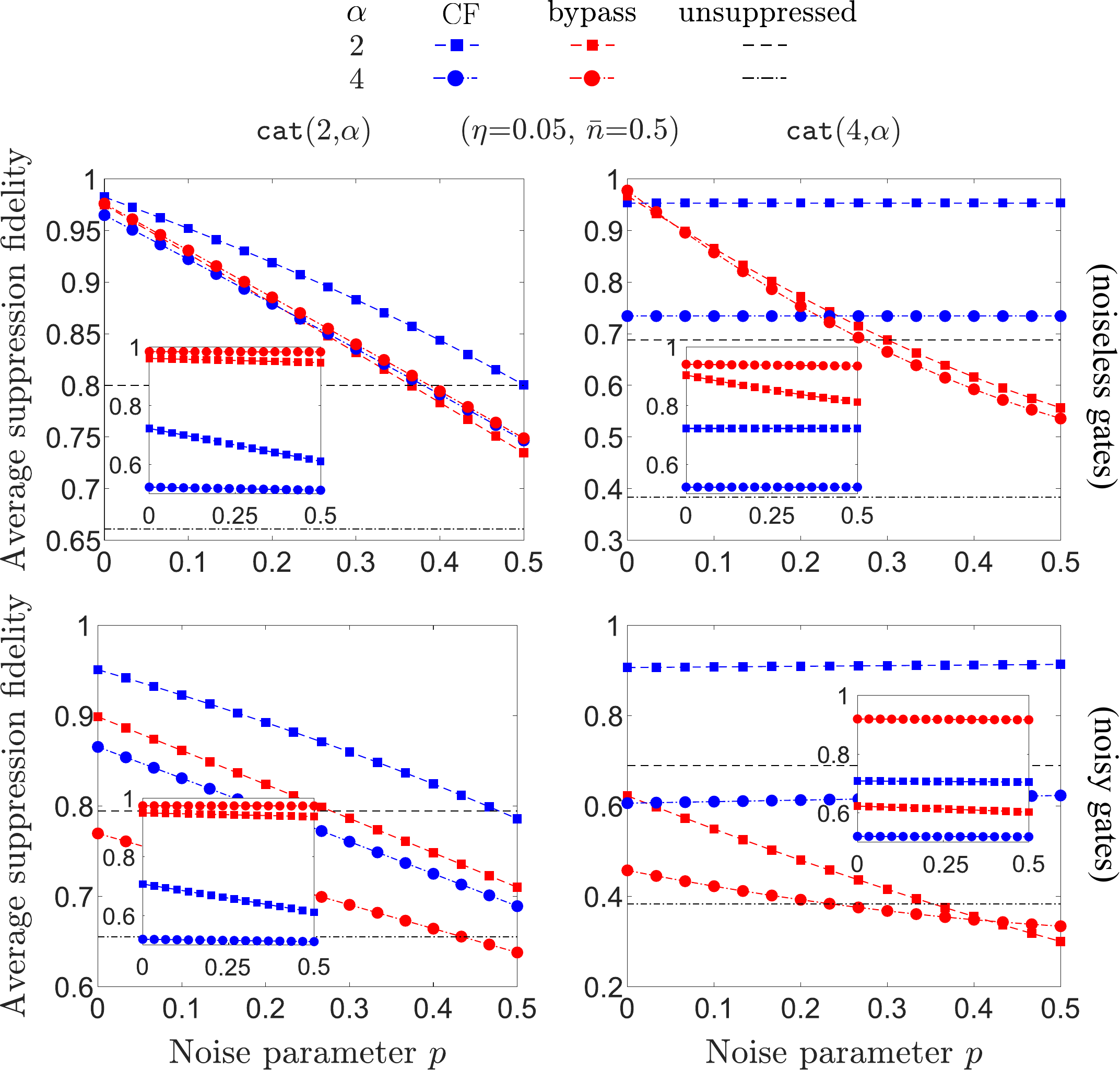}
			\caption{\label{fig:compare_bypass}
				Comparison between using conditional Fouriers and the conditional displacements prescribed by the ``bypass" protocol of Ref.~\cite{Park2022:Slowing} for two- and four-component cat codes. The latter uses a total of four conditional gates for the  $\texttt{cat}(2,\alpha)$ codes. An additional DV ancilla, along with four more conditional displacements 
				and two bosonic rotations by $\pi/4$, 
				were used in ``bypass'' for~$\texttt{cat}(4,\alpha)$. Although the opposite-parity codes are not robust to the DV damping noise in our protocol, the results show that the performance is better for moderately high $|\alpha|\sim 2$ and more robust to unknown DV damping noise. Moreover, our CF-only interferometer uses at most two conditional gates with a \mbox{single~ancilla}. The differences in the number of gates become important when the conditional gates are noisy. Here, all noisy entangling gates have additional 1\% loss and composite DV damping rates. Insets show average success~probability.}
		\end{figure}

		Figures~\ref{fig:likevsopp} and~\ref{fig:PQPcondrot_damp} demonstrated that like-parity codes are impervious to DV damping noise because no state transfer from CV to DV occurs at any point, unlike the bypass strategy. We now cement the distinction with simple examples of opposite-parity codes (the two-component cat codes, $\texttt{cat}(2,\alpha)$ in~Fig.~\ref{fig:compare_bypass} where codes still outperform complete quantum state transferring, where the information suffers from the DV noise without suppression. Therefore, our scheme is preferable to ``bypass" protocols in realistic regimes where bosonic codes have moderately large average photon numbers~($\sim4$) and DV ancillae are affected by a potentially large, uncalibrated damping~noise.        

\begin{figure}[t!]
			\centering
			\includegraphics[width=\columnwidth]{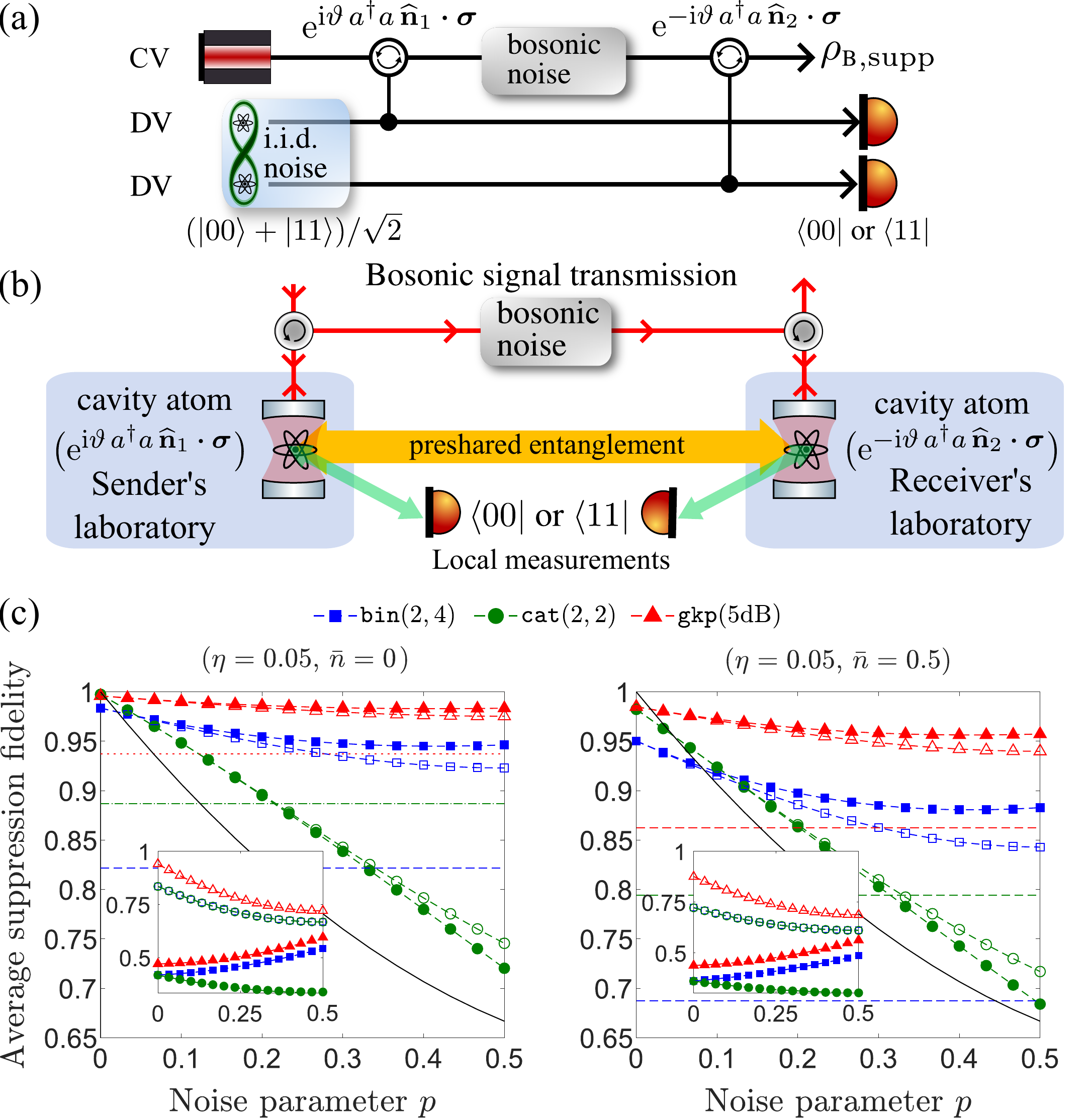}
			\caption{\label{fig:comm} 
				The like parity codes are also resilient to the composite amplitude and phase damping ancilla noise in quantum communication.
				(a) Circuit for bosonic signal transmission with preshared DV entanglement and classical communications as a resource. (b) An optical circuit to implement the protocol for quantum signal transmission using two remote cavity atom systems. (c) The average performance of various encodings under the noise suppression protocol for both photon loss of rate $\eta=0.05$ and thermal noise of the same rate and $\bar{n}=0.5$. The unmarked dashed, dot-dashed, and dotted lines refer to the average unsuppressed fidelities for the binomial, cat, and GKP codes, while the solid one is the average DV teleportation fidelity. Insets show average success~probability. The filled and unfilled markers refer to heralding the bosonic output on the ancillary measurement outcome~$00$, and both outcomes $00$ and $11$,~respectively.}
		\end{figure}

		\emph{Noise suppression for quantum communication.---}We have only looked at localized suppression unitaries but controlled rotations for traveling waves could be best exploited in the nonlocal applications, such as for transmitting qumode states over a \mbox{noisy~channel}~\cite{Munro2012:Quantum, Li2023:Memoryless, Li2024:Performance}. Preshared entanglement \mbox{between} atoms inside remote cavities and classical communications can be used as resources~\cite{Ritter2012:Elementary} as shown in~\mbox{Figs.~\ref{fig:comm}~(a) and (b)}. For perfect ancillae, $\overline{\mathcal{F}}_\mathrm{supp}$ is as in Eq.~(\ref{eq:Fsupp}), and the success rate \emph{coincides} with Eq.~\eqref{eq:psucc}.
		
		Such setups enable stationary systems to be utilized as long-lived memories for resource states and local gate operations, while traveling waves are used for communication and correction. Moreover, all the measurements are fixed projections without feedforwarding, unlike for teleportation~\cite{Munro2012:Quantum}. 
		
		Results for various bosonic codes in Fig.~\ref{fig:comm} show the advantage of our protocol in the presence of composite DV damping noise on the resource \mbox{Bell~state}.
		The average DV-only teleportation fidelity assuming that only composite amplitude and phase damping noise acts on the shared Bell state is given by $\overline{\mathcal{F}}_\mathrm{tele}=1-p+2p^2/3$ (Sec.~{E} of SM~\cite{SM}) and the figure reveals that teleportation requires high-quality DV Bell states, whereas our scheme is resilient to damping~noise. Moreover, the DV-only protocol does not support further error correction.
		
		The performance in terms of the average success probability of this protocol is given by (Sec.~{D} of SM~\cite{SM}),
		\begin{equation}
			\overline{p}_\mathrm{succ}=\frac{1}{2}+\frac{1-2p(1-p)}{2(2G-1)}\tr{\SL
				\left(\frac{1-2\mu G}{2G-1}\right)^{\AdA}},
			\label{eq:psucc_00_or_11}
		\end{equation}
		when the output is heralded on \emph{both} $00$ and $11$~outcomes. With damping noise, the average fidelity of the even-parity codes may be slightly enhanced by only heralding on~$00$ as shown by the filled markers in Fig.~\ref{fig:comm}, at a reduced success~probability.
		
		\emph{Conclusion and outlook.---}We analyzed a noise suppression protocol that targets photon-loss and thermal-noise corruption of single-mode bosonic systems or~qumodes. It uses the hybrid entangling operations between the qumode and a single qubit ancilla, specifically the conditional Fourier gates sandwiching the noise along with a final nondeterministic projection of the DV ancilla onto its original~state. 
		
		Potential applications lie in quantum computing and  communication, as it addresses photon losses and thermal noise, which are the dominant noise sources for~qumodes. While linear-optical mitigation methods using probabilistic error cancellation inevitably require measurements that reduce the bosonic-mode quantum state to classical numbers and, at the same time, suffer from large sampling costs,
		our noise suppression method retains the superposition state for further \emph{quantum} information processing, with a high success~rate. 
		
		In the context of quantum computing with bosonic codes encoding qubit~information, we show that codes with orthogonal codewords of the same parity (whether odd or even) are completely impervious to additional ancilla damping noise than those having opposite-parity codewords. Here, we present a simple setup with a maximum of two conditional Fourier gates, making it experimentally feasible. In the future, suppression techniques with multiple qubit ancillae in the presence of correlated DV noise could be~developed.
		
		Further investigation of the error correction performances of bosonic codes when supplemented with feasible noise suppression protocols, such as the one presented here, is the natural route to~take. Bosonic codes that are easier to realize, but less capable in terms of error-correcting performance, may become competitive candidates after such noise-robust hybrid suppression~protocols. This trade-off needs to be explored on a case-by-case basis, depending on specific experimental constraints. Systematic studies on multi-control hybrid rotation gates on additional qubit ancillae are also crucial future goals to probe the performance of hybrid suppression~schemes.
		
		\emph{Acknowledgments.---}The authors thank M.S. Kim, Y. Kim, K. Park, R. Filip, and S. Bose for insightful discussions. This work was supported by the Korean government [Ministry of Science and ICT (MSIT)], the NRF grants funded by the Korea government (MSIT) (No. RS-2023-00237959, No. RS-2024-00413957, No. RS-2024-00438415, No. RS-2025-02219034 and No. RS-2023-NR076733), the Institute of Information \& Communications Technology Planning \& Evaluation (IITP) grant funded by the Korea government (MSIT) (IITP-2025-RS-2020-II201606 and IITP-2025-RS-2024-00437191), and the Institute of Applied Physics at Seoul National University.

		\onecolumngrid
		\begin{center}
			\large\textsc{Supplemental Material}
		\end{center}
		\vspace{-10ex}
		\appendix
		
		\section{Thermal noise suppression with conditional Fourier gates: average fidelity}
		
		Thermal noise of error rate $\etaT$ is characterized by a quantum-limited amplifier of gain $G:=1+\etaT\,\bar{n}$ following a photon-loss channel with rate $\mu:=1-(1-\etal)/G$ (Eq.~\eqref{eq:noise})~\cite{Ivan2011:Operator-sum,Gagatsos2017:Bounding}.
		Consider one pair of Kraus operators corresponding to $l$-photon losses 
		\begin{equation}
			A_l:=\sqrt{\frac{\mu^l}{l!}}\sqrt{1-\mu}^{\frac{\AdA}{2}}a^l
			\label{eq:lossK}
		\end{equation} 
		and $k$-excitations~\cite{Ivan2011:Operator-sum, Gagatsos2017:Bounding}
		\begin{equation}
			B_k:=\sqrt{\frac{(1-G^{-1})^k}{k!\, G}}\,a^{\dagger k}G^{-\frac{\AdA}{2}}.
			\label{eq:ampK}
		\end{equation}
		
		Under the action of the suppression unitary as described in the main text,
		\begin{equation}
			U_s=\E{\I\vartheta \AdA \nhat\bm{\cdot}\bm{\sigma}}
		\end{equation} 
		and it's adjoint, the paired Kraus operators from Eqs.~\eqref{eq:lossK} and \eqref{eq:ampK} transform into
		\begin{align}
			U_s^\dagger B_k A_{l} U_s=&c_{l,k}\sqrt{\frac{1-\mu}{G}}^{\AdA}a^{\dagger k}a^l\E{\I\vartheta(l-k)\nhat\bm{\cdot}\bm{\sigma}},
		\end{align}
		where $c_{l,k}:=\sqrt{\left(\frac{G-1}{1-\mu}\right)^k\frac{\mu^l}{k!\,l!\,G}}$. 
		
		By initializing the ancilla in the $\ket{0}$ and measuring heralding on it being unchanged, we get an unnormalized output $\tilde{\rho}'$ starting from a pure state $\rho$ as,
		\begin{equation}
			\tilde{\rho}'=\sum_{k,l=0}^{\infty}c_{l,k}^2\sqrt{\frac{1-\mu}{G}}^{\AdA}a^{\dagger k}a^l\rho\, a^{\dagger l}a^k\sqrt{\frac{1-\mu}{G}}^{\AdA}|\opinner{0}{\E{\I\vartheta (l-k)\AdA\, \nhat\bm{\cdot}\bm{\sigma}}}{0}|^2.
			\label{eq:supp-state}
		\end{equation}
		
		When $\vartheta=\pi/2$ and $\nhat\perp\khat$, the DV parts give
		\begin{align}
			\opinner{0}{\E{\I\frac{\pi}{2} (l-k) \AdA\, \nhat\bm{\cdot}\bm{\sigma}}}{0}&=\cos((l-k)\pi/2)+\I \opinner{0}{\nhat\bm{\cdot} \bm{\sigma}}{0} \sin((l-k)\pi/2)\nonumber\\
			&=\cos((l-k)\pi/2)\nonumber\\
			\therefore\,|\opinner{0}{\E{\I\frac{\pi}{2} (l-k) \AdA\, \nhat\bm{\cdot}\bm{\sigma}}}{0}|^2&=\delta_{\mathrm{even}\,(l-k)}.
			\label{eq:odd-suppressed}
		\end{align}
		This suppresses noise events that lead to a change in the photon number parity, \emph{without} conventional error correction.
		
		For small error rates $\etaT$, we additionally have the coefficients
		\begin{align}
			c^2_{0,0}&\cong1-\etaT\bar{n}+\etaT^2\bar{n}^2,\!\!\!\!\!\!\!\!\!\!\!\!\!\!\!\!\!\!\!\!\!\!\!\!\!\!\!\!\!\!\!\!\!\!\!\!\!\!\!\!\!\!\!\!\!\!\!\!\!\!\!\!\!\!\!\!\!\!\!\!\!\!\!\!\!\!\!\!\!\!\!\!\!\!\!\!\!\!\!\!\!\!\!\!\!\!\!\!\!\!\!\!\!\!\!\!\!\!\!\!\!\!&&c^2_{1,1}\cong\etaT^2\bar{n}(1+\bar{n}),\nonumber\\
			c^2_{1,0}&\cong\etaT(1+\bar{n})-2\bar{n}\etaT^2(1+\bar{n}),\!\!\!\!\!\!\!\!\!\!\!\!\!\!\!\!\!\!\!\!\!\!\!\!\!\!\!\!\!\!\!\!\!\!\!\!\!\!\!\!\!\!\!\!\!\!\!\!\!\!\!\!\!\!\!\!\!\!\!\!\!\!\!\!\!\!\!\!\!\!\!\!\!\!\!\!\!\!\!\!\!\!\!\!\!\!\!\!\!\!\!\!\!\!\!\!\!\!\!\!\!\!\!\!\!\!\!\!\!\!\!\!\!\!\!\!\!\!\!\!\!\!\!\!\!\!&&c^2_{0,1}\cong\etaT\bar{n}+\etaT^2\bar{n},\nonumber\\
			c^2_{0,2}&\cong\frac{\etaT^2}{2}(1+\bar{n})^2,\!\!\!\!\!\!\!\!\!\!\!\!\!\!\!\!\!\!\!\!\!\!\!\!\!\!\!\!\!\!\!\!\!\!\!\!\!\!\!\!\!\!\! &&c^2_{2,0}\cong\frac{\etaT^2}{2}\bar{n}^2,
		\end{align}
		and the \emph{rescaling} operator
		\begin{align}
			&\sqrt{\frac{1-\mu}{G}}^{\AdA}\cong1-\etaT T_1+\etaT^2 T_2,\nonumber\\
			&T_1:=\left(\bar{n}+\frac{1}{2}\right){\AdA},\nonumber\\
			&T_2:=\frac{1}{8}\left[(\AdA)^2(1+2\bar{n})^2+2(2\bar{n}^2-1)\AdA\right].
		\end{align}
		
		Using these approximations while restricting to a maximum of two noisy jumps, together with the observation from Eq.~(\ref{eq:odd-suppressed}), for a noiseless DV ancilla, we obtain that
		\begin{align}
			\tilde{\rho}'=\rho&-\etaT(\bar{n}\rho+\{T_1,\rho\})\nonumber\\
			&+\etaT^2\bigg[\bar{n}^2\rho+\bar{n}\{T_1,\rho\}+T_1\rho\,T_1+\{T_2,\rho\}+\bar{n}(1+\bar{n})\AdA\rho\,\AdA+\frac{(1+\bar{n})^2}{2}a^2\rho\,a^{\dagger 2}+\frac{\bar{n}^2}{2}a^{\dagger 2}\rho\,a^{2}\bigg].
		\end{align}
		
		The normalization of this state leads to the success probability
		\begin{align}
			p_\mathrm{succ}&\cong1-\etaT[\bar{n}+(1+2\bar{n}\MEAN{\AdA})]+\etaT^2\bigg[(1+2\bar{n})^2\MEAN{(\AdA)^2}+(4\bar{n}^2-1)\MEAN{\AdA}+2\bar{n}^2\bigg],
		\end{align}
		employing $\MEAN{\cdot}$ for the expectation values in the state $\rho$, and 
		\begin{align}
			a^{\dagger 2}a^2=&\AdA (\AdA-1),\nonumber\\
			a^{2}a^{\dagger 2}=&(\AdA+1)(\AdA+2), 
		\end{align}
		using the commutators of the bosonic ladder operators.
		
		Now, using
		\begin{align}
			\frac{1}{1-ax+b x^2}\cong1+ax+(a^2-b)x^2
		\end{align}
		for $x\ll 1$, we have,
		\begin{equation}
			p^{-1}_{\mathrm{succ}}\cong1+\etaT[\bar{n}(1+2\bar{n}\MEAN{\AdA})]+\etaT^2\bigg[(1+2\bar{n})^2(\MEAN{\AdA}^2-\MEAN{(\AdA)^2})+(1+2\bar{n})\MEAN{\AdA}-\bar{n}^2\bigg].
		\end{equation}
		
		The normalized output state is therefore,
		\begin{align}
			&\rho'=\rho-\etaT\bigg[\frac{1+2\bar{n}}{2}(\AdA \rho+\rho \AdA)-(1+2\bar{n})\MEAN{\AdA}\rho\bigg]\nonumber\\
			&-\etaT^2\bigg[\frac{1+2\bar{n}}{2}(\AdA\rho+\rho\AdA)+\bar{n}\rho\bigg]\big[(1+2\bar{n})\MEAN{\AdA}+\bar{n}\big]\nonumber\\
			&+\etaT^2\big[(1+2\bar{n})^2(\MEAN{\AdA}^2-\MEAN{(\AdA)^2}+(1+2\bar{n})\MEAN{\AdA}-\bar{n}^2
			\big]\nonumber\\
			&+\etaT^2\bigg[\frac{(1+2\bar{n})^2}{4}\AdA\rho\AdA\nonumber\\
			&+\bigg(\frac{\bar{n}(1+3\bar{n})}{2}-\frac{1}{4}\bigg)(\AdA\rho+\rho\AdA)\nonumber\\
			&+\frac{(1+2\bar{n})^2}{8}[(\AdA)^2\rho+\rho(\AdA)^2]+\bar{n}\rho\nonumber\\
			&+\bar{n}(1+\bar{n})\AdA\rho\AdA+\frac{\bar{n}^2}{2}a^{\dagger 2}\rho\,a^2+\frac{(1+\bar{n})^2}{2}a^2\rho\,a^{\dagger 2}\bigg].
		\end{align}
		
		The fidelity is given by $\tr{\rho \rho'}$ for the pure state $\rho$, and it is immediately clear that $\mathcal{O}(\etaT)$, term vanishes and it becomes
		\begin{align}
			\mathcal{F}_{\mathrm{supp}}\cong1-\etaT^2\bigg[&\bar{n}^2+3\left(\bar{n}+\frac{1}{2}\right)^2\MEAN{(\AdA)^2}+\left(\bar{n}^2-\bar{n}-\frac{1}{2}\right)\MEAN{\AdA}\nonumber\\
			&-\left(\frac{1}{4}+2\bar{n}(1+\bar{n})\right)\tr{\rho\,\AdA\rho\,\AdA}-\left(\frac{1}{2}+\bar{n}+\bar{n}^2\right)\tr{\rho\,a^2\rho\,a^{\dagger 2}}
			\bigg].
		\end{align}
		
		We now use various identities of the averages over the Haar measure of unitaries for a single qubit to find the fidelity averaged over the heralded states as
		\begin{align}
			\overline{\mathcal{F}}_\mathrm{supp}\cong&\, 1-\etaT^2\bigg\{\bar{n}^2+3\,\left(\bar{n}+\frac{1}{2}\right)^2\,\tr{\SL\,(a^\dag a)^2}+\left(\bar{n}^2-\bar{n}-\frac{1}{2}\right)\tr{\SL\,a^\dag a}\nonumber\\
			&\,-\left[\frac{1}{6}+\frac{4}{3}(\bar{n}^2+\bar{n})\right]\left(\tr{\left(\SL\,a^\dag a\right)^2}+\tr{\SL\,a^\dag a}^2\right)-\left[\frac{1}{3}+\frac{2}{3}\left(\bar{n}^2+\bar{n}\right)\right]\left(\tr{\SL a^2\,\SL a^{\dag2}}+\left|\tr{\SL a^2}\right|^2\right)\bigg\},
		\end{align}
		
		{\allowdisplaybreaks
			\section{Thermal noise suppression with conditional Fourier gates: average success probability}
			
			Recall that the unnormalized state at the output is 
			\begin{align}
				\tilde{\rho}'=\sum_{k,l\geq0}&c_{l,k}^2\sqrt{\frac{1-\mu}{G}}^{\AdA}a^{\dagger k}a^l\rho\, a^{\dagger l}a^k\sqrt{\frac{1-\mu}{G}}^{\AdA}\delta_{\mathrm{even}\,(l-k)}.
			\end{align}
			
			The trace of this state gives the success probability
			\begin{align}
				&\tr{\tilde{\rho}^\prime}=\mathop{\mathrm{tr}}\bigg\{\rho\sum_{l,k\geq0} c_{l,k}^2 \,a^{\dagger l}a^k\left(\frac{1-\mu}{G}\right)^{\AdA}a^{\dagger k}a^l\delta_{\mathrm{even}\,(l-k)}\bigg\}\nonumber\\
				&=\mathop{\mathrm{tr}}\bigg\{\rho\left(\frac{1-\mu}{G}\right)^{\AdA}\frac{1}{G}\sum_{l\geq0}\frac{1}{l!}\left(\frac{\mu G}{1-\mu}\right)^l a^{\dagger l}\left[\sum_{k\geq 0}\left(1-\frac{1}{G}\right)^k\frac{1}{k!}a^ka^{\dagger k}\delta_{\mathrm{even}\,(l-k)}\right]a^l 
				\bigg\}\nonumber\\
				&=\frac{1}{G}\mathop{\mathrm{tr}}\bigg\{\rho\left(\frac{1-\mu}{G}\right)^{\AdA}\bigg[\sum_{\mathrm{even}\,l}\frac{y^l}{l!}
				a^{\dagger l}\vdots\sum_{\mathrm{even}\,k}\frac{z^k}{k!}(\AdA)^k\vdots a^l+\sum_{\mathrm{odd}\,l}\frac{y^l}{l!}
				a^{\dagger l}\vdots\sum_{\mathrm{odd}\,k}\frac{z^k}{k!}(\AdA)^k\vdots a^l\bigg]\bigg\}\nonumber\\
				&=\frac{1}{2G}\mathop{\mathrm{tr}}\bigg\{\rho\left(\frac{1-\mu}{G}\right)^{\AdA}\bigg[\sum_{\mathrm{even}\,l}\frac{y^l}{l!}
				a^{\dagger l}\vdots\E{z\AdA}+\E{-z\AdA}\vdots a^l+\sum_{\mathrm{odd}\,l}\frac{y^l}{l!}
				a^{\dagger l}\vdots\E{z\AdA}-\E{-z\AdA}\vdots a^l\bigg]\bigg\}\\
				&=\frac{1}{2G}\mathop{\mathrm{tr}}\bigg\{\rho\left(\frac{1-\mu}{G}\right)^{\AdA}\bigg[\sum_{\mathrm{even}\,l}\frac{y^l}{l!}
				a^{\dagger l}\left(\E{-\ln(1-z)}\E{-\ln(1-z)\AdA}+\E{-\ln(1+z)}\E{-\ln(1+z)\AdA}\right)  a^l\nonumber\\
				&{\hskip10em\relax}+\sum_{\mathrm{odd}\,l}\frac{y^l}{l!}
				a^{\dagger l}(\E{-\ln(1-z)}\E{-\ln(1-z)\AdA}-\E{-\ln(1+z)}\E{-\ln(1+z)\AdA}) a^l\bigg]\bigg\}\nonumber\\
				&=\frac{1}{2G}\mathop{\mathrm{tr}}\bigg\{\rho\left(\frac{1-\mu}{G}\right)^{\AdA}\bigg[\sum_{\mathrm{even}\,l}\frac{y^l}{l!}
				a^{\dagger l}\left(\frac{:\E{\AdA(\E{-\ln(1-z)}-1)}:}{1-z}+\frac{:\E{\AdA(\E{-\ln(1+z)}-1)}:}{1+z}\right)  a^l\nonumber\\
				&{\hskip10em\relax}+\sum_{\mathrm{odd}\,l}\frac{y^l}{l!}
				a^{\dagger l}\left(\frac{:\E{\AdA(\E{-\ln(1-z)}-1)}:}{1-z}-\frac{:\E{\AdA(\E{-\ln(1+z)}-1)}:}{1+z}\right) a^l\bigg]\bigg\}\nonumber\\
				&=\frac{1}{2G}\mathop{\mathrm{tr}}\bigg\{\rho\left(\frac{1-\mu}{G}\right)^{\AdA}\bigg[\sum_{\mathrm{even}\,l}\frac{y^l}{l!}
				a^{\dagger l}\left(\frac{:\E{\frac{z}{1-z}\AdA}:}{1-z}+\frac{:\E{\frac{-z}{1+z}\AdA}:}{1+z}\right)a^l+\sum_{\mathrm{odd}\,l}\frac{y^l}{l!}
				a^{\dagger l}\left(\frac{:\E{\frac{z}{1-z}\AdA}:}{1-z}-\frac{:\E{\frac{-z}{1+z}\AdA}:}{1+z}\right) a^l\bigg]\bigg\}\nonumber\\
				&=\frac{1}{2G}\mathop{\mathrm{tr}}\bigg\{\rho\left(\frac{1-\mu}{G}\right)^{\AdA}\bigg[\sum_{\mathrm{even}\,l}\frac{y^l}{l!}
				\left(\frac{:(\AdA)^l\E{\frac{z}{1-z}\AdA}:}{1-z}+\frac{:(\AdA)^l\E{\frac{-z}{1+z}\AdA}:}{1+z}\right)\nonumber\\
				&{\hskip10em\relax}+\sum_{\mathrm{odd}\,l}\frac{y^l}{l!}
				\left(\frac{:(\AdA)^l\E{\frac{z}{1-z}\AdA}:}{1-z}-\frac{:(\AdA)^l\E{\frac{-z}{1+z}\AdA}:}{1+z}\right) \bigg]\bigg\}\nonumber\\
				&=\frac{1}{4G}\mathop{\mathrm{tr}}\bigg\{\rho\left(\frac{1-\mu}{G}\right)^{\AdA}\bigg[
				:\left(\E{y\AdA}+\E{-y\AdA}\right)\frac{\E{\frac{z}{1-z}\AdA}}{1-z}:+:\left(\E{y\AdA}+\E{-y\AdA}\right)\frac{\E{-\frac{z}{1+z}\AdA}}{1+z}:\nonumber\\
				&{\hskip10em\relax}+:\left(\E{y\AdA}-\E{-y\AdA}\right)\frac{\E{\frac{z}{1-z}\AdA}}{1-z}:-:\left(\E{y\AdA}-\E{-y\AdA}\right)\frac{\E{-\frac{z}{1+z}\AdA}}{1+z}:\bigg]\bigg\}\nonumber\\
				&=\frac{1}{4G}\mathop{\mathrm{tr}}\bigg\{\rho\left(\frac{1-\mu}{G}\right)^{\AdA}\bigg[
				\frac{\left(1+y+\frac{z}{1-z}\right)^{\AdA}}{1-z}+\frac{\left(1-y+\frac{z}{1-z}\right)^{\AdA}}{1-z}+\frac{\left(1+y-\frac{z}{1+z}\right)^{\AdA}}{1+z}+\frac{\left(-y-\frac{z}{1+z}\right)^{\AdA}}{1+z}\nonumber\\
				&{\hskip10em\relax}+\frac{\left(1+y+\frac{z}{1-z}\right)^{\AdA}}{1-z}-\frac{\left(1-y+\frac{z}{1-z}\right)^{\AdA}}{1-z}-\frac{\left(1+y-\frac{z}{1+z}\right)^{\AdA}}{1+z}+\frac{\left(1-y-\frac{z}{1+z}\right)^{\AdA}}{1+z}\bigg]\bigg\}\nonumber\\
				&=\frac{1}{2G}\mathop{\mathrm{tr}}\bigg\{\rho\left(\frac{1-\mu}{G}\right)^{\AdA}\bigg[
				\frac{\left(1+y+\frac{z}{1-z}\right)^{\AdA}}{1-z}+\frac{\bigg(1-y-\frac{z}{1+z}\bigg)^{\AdA}}{1+z}\bigg]\bigg\}\nonumber\\
				&=\frac{1}{2}+\frac{1}{2(2G-1)}\mathop{\mathrm{tr}}\bigg\{\rho
				\bigg(\frac{1-2\mu G}{2G-1}\bigg)^{\AdA}\bigg\}.
			\end{align}
			where we defined $y:={\mu G}/({1-\mu})$ and $z:=1-{1}/{G}$. Here $:\cdot:$ and $\vdots\cdot\vdots$ are the normal and the antinormal orderings, respectively. 
			
			Averaging over the encoded states leads to,
			\begin{equation}
				\overline{p}_{\mathrm{succ}}=\frac{1}{2}+\frac{1}{2(2G-1)}\mathop{\mathrm{tr}}\bigg\{\SL
				\bigg(\frac{1-2\mu G}{2G-1}\bigg)^{\AdA}\bigg\}
				\label{eq:psucc_1dv}
			\end{equation}
			
			\section{Average fidelity under thermal noise without suppression}
			
			Without any noise suppression, Eq.~\eqref{eq:supp-state} doesn't posses the final DV term and therefore, the pairings $(k,l)=(1,0)$ and $(0,1)$ do not vanish.
			
			Consequently, the (normalized) state is
			\begin{equation}
				\rho'\cong\rho-\eta\args{\bar{n}\rho+\argc{\argp{\bar{n}+\frac{1}{2}}\AdA,\rho} -\bar{n}a^\dagger\,\rho\,a -(1+\bar{n})a\,\rho\,a^\dagger},
			\end{equation}
			giving the fidelity,
			\begin{align}
				\tr{\rho \rho'}\cong&1-\eta\args{\bar{n}+(1+2\bar{n})\MEAN{\AdA}-\bar{n}\tr{\rho\,a^\dagger\,\rho\,a}-(1+\bar{n})\tr{\rho\,a\,\rho\,a^\dagger}}.
			\end{align}
			
			The above equation is averaged to
			\begin{equation}
				\overline{\mathcal{F}}_{\mathrm{unsupp}}\cong1-\eta\args{\bar{n}+(1+2\bar{n})\tr{\SL\,\AdA} -\frac{2\bar{n}}{3}\argp{\tr{\SL\,a^\dagger\,\SL\,a}+|\tr{\SL\,a}|^2}-\frac{2(1+\bar{n})}{3}\argp{\tr{\SL\,a\SL\,a^\dagger}+|\tr{\SL\,a}|^2}},
			\end{equation}
			where we now see the undesirable linear scaling with $\eta$. Moreover, note that the final two terms in the parentheses vanish for the like parity codes, suggesting they are affected more severely than their opposite parity counterparts with the identical mean photon number ($\tr{\SL\,\AdA}$).
			
			\section{Success probability of quantum communication setup under thermal noise and ancilla damping noise}
			
			The setup for bosonic signal transmission requires a preshared entanglement stored in the remote atoms in the form of a Bell state $\ket{\Phi^+}=\big(\ket{00}+\ket{11}\big)/\sqrt{2}$. Here, we consider the effect of the composite DV damping on the Bell state ancilla. The iid damping noise channel over two qubits can be constructed using  Eq.~\eqref{eq:damp} on the various terms of the density operator $\rho_{\Phi^+}=\ketbra{\Phi^+}{\Phi^+}$ as
			\begin{align}
				\mathcal{N}_{\text{damp}}^{\otimes 2}[\ketbra{00}{00}]=&\frac{(c_++c_-+2p)^2}{16}\ketbra{00}{00}=(1+p^2)\ketbra{00}{00},\nonumber\\
				\mathcal{N}_{\text{damp}}^{\otimes 2}[\ketbra{11}{11}]=&\frac{(c_++c_--2p)^2}{16}\ketbra{00}{00}+p^2\ketbra{11}{11}+\frac{p(c_++c_--2p)}{4}\big(\ketbra{01}{01}+\ketbra{10}{10}\big),\nonumber\\
				=&(1-p)^2\ketbra{00}{00}+p^2\ketbra{11}{11}+p(1-p)\big(\ketbra{01}{01}+\ketbra{10}{10}\big),\nonumber\\
				\mathcal{N}_{\text{damp}}^{\otimes 2}[\ket{00}\!\!\bra{11}]=&\frac{(c_+-c_-)^2}{16}\ket{00}\!\!\bra{11}=(1-p)^2\ket{00}\!\!\bra{11},\nonumber\\
				\mathcal{N}_{\text{damp}}^{\otimes 2}[\ket{11}\!\!\bra{00}]=&\frac{(c_+-c_-)^2}{16}\ket{11}\!\!\bra{00}=(1-p)^2\ket{11}\!\!\bra{00},
			\end{align}
			which leads to,
			\begin{align}
				\mathcal{N}_{\text{damp}}^{\otimes 2}[\rho_{\Phi^+}]=&\frac{1}{2}\bigg\{
				(1+p^2)\ketbra{00}{00}+(1-p)^2\ketbra{11}{11}+p(1-p)\big(\ketbra{01}{01}+\ketbra{10}{10}\big)
				+(1-p)^2\,\big(\ket{00}\!\!\bra{11}+\ket{11}\!\!\bra{00}\big)
				\bigg\}.
				\label{eq:noisy-Bell}
			\end{align}
			
			Due to the linearity of transformations, we find the relevant inner products of the form $K_{s}(\mu,\nu,l,k):=\opinner{}{U_s^{(2)} B_kA_l U_s^{(1)}}{\mu\nu}$ where $\bra{}$ denotes the conditioned outcomes (either $\bra{00}$ or $\bra{11}$) and $\ket{\mu\nu}$ is ket representing the state of two DV ancilla with $\mu,\nu=$ 0 or 1, $A_l$ and $B_k$ are the $l$th and the $k$th Kraus operators of loss and quantum-limited amplification channels respectively, and $U_s^{(1)}:=\exp\big(\I \frac{\pi}{2}\AdA\, \sigma_1^{(1)}\big)$ and $U_s^{(2)}:=\exp\big(-\I \frac{\pi}{2}\AdA\, \sigma_1^{(2)}\big)$ are the unitaries implemented in the sender's and receiver's laboratory respectively. These inner products can be viewed as suppressed Kraus operators, which may not be complete and trace-preserving.
			
			Now, we follow the procedure similar to obtaining Eq.~(\ref{eq:psucc_1dv}). First, by dropping the parameters in the notation for simplicity, we have the suppressed Kraus operators,
			\begin{align}
				&K_s=c_{l,k}\sqrt{\frac{1-\mu}{G}}^{\AdA}\,\opinner{}{\E{-\I \frac{\pi}{2}\AdA \sigma_1^{(2)}} a^{\dagger k}a^l\E{\I \frac{\pi}{2}\AdA \sigma_1^{(1)}}}{\mu\nu}\nonumber\\
				&=c_{l,k}\sqrt{\frac{1-\mu}{G}}^{\AdA}a^{\dagger k} a^l\,\big[\cos\argp{\frac{\pi}{2}(\AdA+k-l)}\cos\left(\frac{\pi}{2}\AdA\right)\inner{}{\mu\nu}+\I\cos\left(\frac{\pi}{2}(\AdA+k-l)\right)\sin\left(\frac{\pi}{2}\AdA\right)\opinner{}{\sigma_1^{(1)}}{\mu\nu}\nonumber\\
				&{\hskip8em\relax}-\I\sin\argp{\frac{\pi}{2}(\AdA+k-l)}\cos\left(\frac{\pi}{2}\AdA\right)\opinner{}{\sigma_1^{(2)}}{\mu\nu}+\sin\argp{\frac{\pi}{2}(\AdA+k-l)}\sin\left(\frac{\pi}{2}\AdA\right)\opinner{}{\sigma_1^{2}\sigma_1^{(1)}}{\mu\nu}\big].
				\label{eq:suppK}
			\end{align}
			
			Using these, we get the unnormalized density operators,
			\begin{align}
				\tilde{\rho}^\prime(00)=&\frac{1}{2}\sum_{l,k\geq0}c_{l,k}^2\sqrt{\frac{1-\mu}{G}}^{\AdA}a^{\dagger k}a^l\bigg\{(1+p^2)\,\cos\argp{\frac{\pi}{2}(\AdA+k-l)}\cos\left(\frac{\pi}{2}\AdA\right)\rho\cos\left(\frac{\pi}{2}\AdA\right)\cos\argp{\frac{\pi}{2}(\AdA+k-l)} \nonumber\\
				&+(1-p)^2\, \sin\argp{\frac{\pi}{2}(\AdA+k-l)}\sin\left(\frac{\pi}{2}\AdA\right)\rho\sin\left(\frac{\pi}{2}\AdA\right)\sin\argp{\frac{\pi}{2}(\AdA+k-l)} \nonumber\\
				&+p(1-p)\,\bigg[ \cos\argp{\frac{\pi}{2}(\AdA+k-l)}\sin\left(\frac{\pi}{2}\AdA\right)\rho\sin\left(\frac{\pi}{2}\AdA\right)\cos\argp{\frac{\pi}{2}(\AdA+k-l)}\nonumber\\
				&{\hskip6em\relax}+\sin\argp{\frac{\pi}{2}(\AdA+k-l)}\cos\left(\frac{\pi}{2}\AdA\right)\rho\cos\left(\frac{\pi}{2}\AdA\right)\sin\argp{\frac{\pi}{2}(\AdA+k-l)}
				\bigg]\nonumber\\
				&+(1-p)^2\,\bigg[ \cos\argp{\frac{\pi}{2}(\AdA+k-l)}\cos\left(\frac{\pi}{2}\AdA\right)\rho\sin\left(\frac{\pi}{2}\AdA\right)\sin\argp{\frac{\pi}{2}(\AdA+k-l)}\nonumber\\
				&{\hskip6em\relax}+\sin\argp{\frac{\pi}{2}(\AdA+k-l)}\sin\left(\frac{\pi}{2}\AdA\right)\rho\cos\left(\frac{\pi}{2}\AdA\right)\cos\argp{\frac{\pi}{2}(\AdA+k-l)}
				\bigg]
				\bigg\}a^{\dagger l}a^k\sqrt{\frac{1-\mu}{G}}^{\AdA}.
			\end{align}
			and similarly,
			\begin{align}
				\tilde{\rho}^\prime(11)=&\frac{1}{2}\sum_{l,k\geq0}c_{l,k}^2\sqrt{\frac{1-\mu}{G}}^{\AdA}a^{\dagger k}a^l\bigg\{(1-p)^2\,\cos\argp{\frac{\pi}{2}(\AdA+k-l)}\cos\left(\frac{\pi}{2}\AdA\right)\rho\cos\left(\frac{\pi}{2}\AdA\right)\cos\argp{\frac{\pi}{2}(\AdA+k-l)} \nonumber\\
				&+(1+p^2)\, \sin\argp{\frac{\pi}{2}(\AdA+k-l)}\sin\left(\frac{\pi}{2}\AdA\right)\rho\sin\left(\frac{\pi}{2}\AdA\right)\sin\argp{\frac{\pi}{2}(\AdA+k-l)} \nonumber\\
				&+p(1-p)\,\bigg[ \cos\argp{\frac{\pi}{2}(\AdA+k-l)}\sin\left(\frac{\pi}{2}\AdA\right)\rho\sin\left(\frac{\pi}{2}\AdA\right)\cos\argp{\frac{\pi}{2}(\AdA+k-l)}\nonumber\\
				&{\hskip6em\relax}+\sin\argp{\frac{\pi}{2}(\AdA+k-l)}\cos\left(\frac{\pi}{2}\AdA\right)\rho\cos\left(\frac{\pi}{2}\AdA\right)\sin\argp{\frac{\pi}{2}(\AdA+k-l)}
				\bigg]\nonumber\\
				&+(1-p)^2\,\bigg[ \cos\argp{\frac{\pi}{2}(\AdA+k-l)}\cos\left(\frac{\pi}{2}\AdA\right)\rho\sin\left(\frac{\pi}{2}\AdA\right)\sin\argp{\frac{\pi}{2}(\AdA+k-l)}\nonumber\\
				&{\hskip6em\relax}+\sin\argp{\frac{\pi}{2}(\AdA+k-l)}\sin\left(\frac{\pi}{2}\AdA\right)\rho\cos\left(\frac{\pi}{2}\AdA\right)\cos\argp{\frac{\pi}{2}(\AdA+k-l)}
				\bigg]
				\bigg\}a^{\dagger l}a^k\sqrt{\frac{1-\mu}{G}}^{\AdA}.
			\end{align}
			for the two outcomes of interest, $00$ and $11$ respectively.
			
			The probabilities of obtaining the state above are their traces. Therefore, for the outcome $00$ we have,
			\begin{align}
				&\,\tr{\tilde{\rho}^\prime(00)}\nonumber\\
				&=\frac{(1+p^2)}{2G}\tr{\cos\left(\frac{\pi}{2}\AdA\right)\rho\cos\left(\frac{\pi}{2}\AdA\right)x^{\AdA}\sum_{l\geq0}\frac{y^l}{l!}\sum_{k\geq0}\frac{z^k}{k!}\cos\argp{\frac{\pi}{2}(\AdA+k-l)}a^{\dagger l}\vdots (\AdA)^k\vdots\,a^l\cos\argp{\frac{\pi}{2}(\AdA+k-l)}}\nonumber\\
				&+\frac{(1-p)^2}{2G}\tr{\sin\left(\frac{\pi}{2}\AdA\right)\rho\sin\left(\frac{\pi}{2}\AdA\right)x^{\AdA}\sum_{l\geq0}\frac{y^l}{l!}\sum_{k\geq0}\frac{z^k}{k!}\sin\argp{\frac{\pi}{2}(\AdA+k-l)}a^{\dagger l}\vdots (\AdA)^k\vdots\,a^l\sin\argp{\frac{\pi}{2}(\AdA+k-l)}}\nonumber\\
				&+\frac{p(1-p)}{2G}\bigg[\tr{\sin\left(\frac{\pi}{2}\AdA\right)\rho\sin\left(\frac{\pi}{2}\AdA\right)x^{\AdA}\sum_{l\geq0}\frac{y^l}{l!}\sum_{k\geq0}\frac{z^k}{k!}\cos\argp{\frac{\pi}{2}(\AdA+k-l)}a^{\dagger l}\vdots (\AdA)^k\vdots\,a^l\cos\argp{\frac{\pi}{2}(\AdA+k-l)}}\nonumber\\
				&{\hskip4em\relax}+\tr{\cos\left(\frac{\pi}{2}\AdA\right)\rho\cos\left(\frac{\pi}{2}\AdA\right)x^{\AdA}\sum_{l\geq0}\frac{y^l}{l!}\sum_{k\geq0}\frac{z^k}{k!}\sin\argp{\frac{\pi}{2}(\AdA+k-l)}a^{\dagger l}\vdots (\AdA)^k\vdots\,a^l\sin\argp{\frac{\pi}{2}(\AdA+k-l)}}\bigg]\nonumber\\
				&+\frac{(1-p)^2}{2G}\bigg[\tr{\sin\left(\frac{\pi}{2}\AdA\right)\rho\cos\left(\frac{\pi}{2}\AdA\right)x^{\AdA}\sum_{l\geq0}\frac{y^l}{l!}\sum_{k\geq0}\frac{z^k}{k!}\cos\argp{\frac{\pi}{2}(\AdA+k-l)}a^{\dagger l}\vdots (\AdA)^k\vdots\,a^l\sin\argp{\frac{\pi}{2}(\AdA+k-l)}}\nonumber\\
				&{\hskip4em\relax}+\tr{\cos\left(\frac{\pi}{2}\AdA\right)\rho\sin\left(\frac{\pi}{2}\AdA\right)x^{\AdA}\sum_{l\geq0}\frac{y^l}{l!}\sum_{k\geq0}\frac{z^k}{k!}\sin\argp{\frac{\pi}{2}(\AdA+k-l)}a^{\dagger l}\vdots (\AdA)^k\vdots\,a^l\cos\argp{\frac{\pi}{2}(\AdA+k-l)}}\bigg]
			\end{align}
			where $x:=(1-\mu)/G$, $y:=\mu\,G/(1-\mu)$, and $z:=1-1/G$. 
			
			The operators are separated from the constant factors using simple trigonometric identities, and $[F(\AdA),a^{\dagger \,l}\vdots(\AdA)^{k}\vdots\,a^l]=0$ for any function of $F(\AdA)$ is used to give,
			\begin{align}
				&\,\tr{\tilde{\rho}^\prime(00)}\nonumber\\
				&=\frac{(1+p^2)}{2G}\mathop{\mathrm{tr}}\Bigg\{\cos\argp{\frac{\pi}{2}\AdA}\rho\cos\argp{\frac{\pi}{2}\AdA}x^{\AdA}\nonumber\\
				&{\hskip4em\relax}\sum_{l\geq0}\frac{y^l}{l!}\sum_{k\geq0}\frac{z^k}{k!}\args{\cos^2\argp{\frac{\pi}{2}\AdA}\delta_{\mathrm{even}\,(k-l)}+\sin^2\argp{\frac{\pi}{2}\AdA}\delta_{\mathrm{odd}\,(k-l)}-\frac{\cancelto{0}{\sin(\pi\AdA)}\cancelto{0}{\sin(\pi(k-l))}}{4}\quad}a^{\dagger \,l}\vdots(\AdA)^{k}\vdots\,a^l\Bigg\}\nonumber\\
				&+\frac{(1-p)^2}{2G}\mathop{\mathrm{tr}}\argc{\sin\argp{\frac{\pi}{2}\AdA}\rho\sin\argp{\frac{\pi}{2}\AdA}x^{\AdA}\sum_{l\geq0}\frac{y^l}{l!}\sum_{k\geq0}\frac{z^k}{k!}\args{\cos^2\argp{\frac{\pi}{2}\AdA}\delta_{\mathrm{odd}\,(k-l)}+\sin^2\argp{\frac{\pi}{2}\AdA}\delta_{\mathrm{even}\,(k-l)}}a^{\dagger \,l}\vdots(\AdA)^{k}\vdots\,a^l}\nonumber\\
				&+\frac{p(1-p)}{2G}\Bigg[\tr{\sin\argp{\frac{\pi}{2}\AdA}\rho\sin\argp{\frac{\pi}{2}\AdA}x^{\AdA}\sum_{l\geq0}\frac{y^l}{l!}\sum_{k\geq0}\frac{z^k}{k!}\args{\cos^2\argp{\frac{\pi}{2}\AdA}\delta_{\mathrm{even}\,(k-l)}+\sin^2\argp{\frac{\pi}{2}\AdA}\delta_{\mathrm{odd}\,(k-l)}}a^{\dagger \,l}\vdots(\AdA)^{k}\vdots\,a^l}\nonumber\\
				&{\hskip4em\relax}+\tr{\cos\left(\frac{\pi}{2}\AdA\right)\rho\cos\argp{\frac{\pi}{2}\AdA}x^{\AdA}\sum_{l\geq0}\frac{y^l}{l!}\sum_{k\geq0}\frac{z^k}{k!}\args{\cos^2\argp{\frac{\pi}{2}\AdA}\delta_{\mathrm{odd}\,(k-l)}+\sin^2\argp{\frac{\pi}{2}\AdA}\delta_{\mathrm{even}\,(k-l)}}a^{\dagger \,l}\vdots(\AdA)^{k}\vdots\,a^l}\Bigg].
			\end{align}
			
			Now, employing the normal and antinormal ordering techniques as in the earlier derivation for Eq.~\eqref{eq:psucc_1dv} we obtain,
			\begin{align}
				&\,\overline{p}_{\mathrm{succ}}(00)=\frac{1}{4}\args{
					1+p+\frac{1-p+2p^2}{2G-1}\tr{\SL \argp{\frac{1-2\mu G}{2G-1}}^{\AdA}}-2p\argp{\tr{\SL \sin^2\!\left(\frac{\pi}{2}\AdA\right)}+\frac{\tr{\SL \sin^2\argp{\frac{\pi}{2}\AdA}\argp{\frac{1-2\mu G}{2G-1}}^{\AdA}}}{2G-1}}
				},
			\end{align}
			and similarly,
			\begin{align}
				&\,\overline{p}_{\mathrm{succ}}(11)=\frac{1}{4}\args{
					1+p+\frac{1-p+2p^2}{2G-1}\tr{\SL \argp{\frac{1-2\mu G}{2G-1}}^{\AdA}}-2p\argp{\tr{\SL \cos^2\!\left(\frac{\pi}{2}\AdA\right)}+\frac{\tr{\SL \cos^2\argp{\frac{\pi}{2}\AdA}\argp{\frac{1-2\mu G}{2G-1}}^{\AdA}}}{2G-1}}
				}.
			\end{align}
			The success rate of accepting both outcomes becomes,
			\begin{equation}
				\,\overline{p}_{\mathrm{succ}}=\frac{1}{2}+\frac{1-2p(1-p)}{2(2G-1)}\tr{\SL\,\argp{\frac{1-2\mu\,G}{2G-1}}^{\AdA}},
			\end{equation}
			which coincides with Eq.~\eqref{eq:psucc_1dv} for $p=0$.
			
			\section{Discrete variable teleportation fidelity with noisy resource}
			
			Here we derive the impact of composite amplitude and phase damping DV noise on the resource entangled state, $\ket{\Phi^+}$ required in the standard DV teleportation protocol. It is one of the four Bell states,
			$\ket{\Phi^\pm}=({\ket{00}\pm\ket{11}})/{\sqrt{2}},$ and $\ket{\Psi^\pm}=({\ket{01}\pm\ket{10}})/{\sqrt{2}}.$
			
			In the first step of the teleportation, a Bell state measurement is performed jointly on the input and one-half of the noisy resource of Eq.~\eqref{eq:noisy-Bell}. 
			The measurement is assumed noiseless positive valued measure, $\mathcal{M}:=\argc{\ketbra{\Phi^+}{\Phi^+},\ketbra{\Phi^-}{\Phi^-},\ketbra{\Psi^+}{\Psi^+},\ketbra{\Psi^-}{\Psi^-}}$ composed of pure Bell states.
			We are genernous with the DV measurement noise for a fairly conservative comparison with our CV-DV hybrid protocol.
			With these preliminary notation the four unnormalized output states, for the DV input $\rho$, are given by,
			\begin{align}
				\tilde{\rho}^\prime(\Phi^\pm)=&\frac{1}{4}\begin{pmatrix}
					\rho_{0,0}(1+p^2)+\rho_{1,1}p(1-p)&\pm\rho_{0,1}(1-p)^2\\
					\pm\rho_{1,0}(1-p)^2&\rho_{1,1}(1-p)^2+\rho_{0,0}p(1-p)
				\end{pmatrix},
			\end{align}
			and
			\begin{align}
				\tilde{\rho}^\prime(\Psi^\pm)=&\frac{1}{4}\begin{pmatrix}
					\rho_{0,0}p(1-p)+\rho_{1,1}(1+p^2)&\pm\rho_{1,0}(1-p)^2\\
					\pm\rho_{0,1}(1-p)^2&\rho_{1,1}p(1-p)+\rho_{0,0}(1-p)^2
				\end{pmatrix}\cdot
			\end{align}
			
			The perfect Pauli corrections $\argc{I,\sigma_z,\sigma_x, \sigma_z\sigma_x}$ are performed on the second half of the noisy resource state corresponding to the four outcomes above, so that the (normalized) teleported state is given by,
			\begin{align}
				\rho_{\mathrm{out}}=&\tilde{\rho}^\prime(\Phi^+)+\sigma_z\tilde{\rho}^\prime(\Phi^-)\sigma_z+\sigma_x\tilde{\rho}^\prime(\Psi^+)\sigma_x+\sigma_z\sigma_x\tilde{\rho}^\prime(\Psi^-)\sigma_x\sigma_z.
			\end{align}
			The fidelity of this mixed state to the pure input is then simply,
			\begin{align}
				\tr{\rho_{\mathrm{in}}\rho_{\mathrm{out}}}=&(1-p)^2+p(1-p)\rho_{0,0}\rho_{1,1}+p(\rho_{0,0}^2+\rho_{1,1}^2)\nonumber\\
				=&1-p+p^2-2p^2\rho_{0,0}(1-\rho_{0,0}),
			\end{align}
			which is averaged to
			\begin{align}
				\overline{\mathcal{F}}_\mathrm{tele}=&1-p+p^2-2p^2\int_0^1\D x\, x(1-x)\nonumber\\
				=&1-p+2p^2/3.
			\end{align}
			
			\section{Error suppression for hybrid entangled states}
			
			Consider hybrid states of the form $\ket{\,\,\,}=\big(\ket{\alpha}\!\!\ket{0}\pm\ket{\beta}\!\!\ket{1}\big)/\sqrt{2}$, which have the density operator
			\begin{equation}
				\rho=\frac{1}{2}\big(\ket{\alpha}\!\!\bra{\alpha}\otimes\ketbra{0}{0}+\ket{\beta}\!\!\bra{\beta}\otimes\ketbra{1}{1}\pm\ket{\alpha}\!\!\bra{\beta}\otimes\ketbra{0}{1}\pm\ket{\beta}\!\!\bra{\alpha}\otimes\ketbra{1}{0}\big).
			\end{equation}
			where $\ket{\alpha}$ and $\ket{\beta}$ are two distinct coherent states.
			
			For photon losses (similar to Eq.~\eqref{eq:suppK}), we have the $l$th Kraus operator sandwiched by the suppression unitaries as
			$
			U_s^\dagger \,K_l\,U_s=\sqrt{\frac{\eta^l}{l!}}\,(1-\eta)^\frac{\AdA}{2}\,a^l\,\E{\I\, l \vartheta\,\nhat\bm{\cdot}\bm{\sigma}}
			$
			where the conditional rotation acts between the qumode of the hybrid state and an \emph{extra} DV qubit, which is the ancilla for the suppression protocol. The dyadic components of the density operator are modified as
			\begin{align}
				\ketbra{\alpha}{\alpha}\otimes\ketbra{0}{0}\otimes\ketbra{0}{0}_{\text{anc}}\mapsto&\sum_{l\geq0}\frac{\eta^l}{l!}\ket{\sqrt{1-\eta}\alpha}\!\ket{0}|\alpha|^{2l}\E{-\eta|\alpha|^2}\bra{0}\!\bra{\sqrt{1-\eta}\alpha}\otimes\E{\I\,l\vartheta\,\nhat\bm{\cdot}\bm{\sigma}}\,\ketbra{0}{0}_{\text{anc}}\,\E{-\I\,l\vartheta\,\nhat\bm{\cdot}\bm{\sigma}},\nonumber\\
				\ketbra{\beta}{\beta}\otimes\ketbra{1}{1}\otimes\ketbra{0}{0}_{\text{anc}}\mapsto&\sum_{l\geq0}\frac{\eta^l}{l!}\ket{\sqrt{1-\eta}\beta}\!\ket{1}|\beta|^{2l}\E{-\eta|\beta|^2}\bra{1}\!\bra{\sqrt{1-\eta}\beta}\otimes\E{\I\,l\vartheta\,\nhat\bm{\cdot}\bm{\sigma}}\,\ketbra{0}{0}_{\text{anc}}\,\E{-\I\,l\vartheta\,\nhat\bm{\cdot}\bm{\sigma}},\nonumber\\
				\ketbra{\alpha}{\beta}\otimes\ketbra{0}{1}\otimes\ketbra{0}{0}_{\text{anc}}\mapsto&\sum_{l\geq0}\frac{\eta^l}{l!}\ket{\sqrt{1-\eta}\alpha}\!\ket{0}(\alpha\beta^*)^l\E{-\frac{\eta}{2}(|\alpha|^2+|\beta|^2)}\bra{1}\!\bra{\sqrt{1-\eta}\beta}\,\E{\I\,l\vartheta\,\nhat\bm{\cdot}\bm{\sigma}}\,\ketbra{0}{0}_{\text{anc}}\,\E{-\I\,l\vartheta\,\nhat\bm{\cdot}\bm{\sigma}},
			\end{align}
			where \mbox{$\ketbra{0}{0}_{\text{anc}}=\big(1+\khat\bm{\cdot}\bm{\sigma}\big)/{2}$} is the initial ancilla state.
			Projecting only the ancilla in the same state gives the familiar $|\opinner{0}{\E{\I\,l\vartheta\,\AdA}}{0}|^2_{\text{anc}}=\delta_{\text{even}(l)}$ for $\vartheta=\pi/2$ and $\nhat \perp \khat$ and hence again responsible for canceling the $\mathcal{O}(\eta)$ terms in the output state. It is clear by observation that similar arguments hold for thermal and Gaussian random displacement noise as discussed earlier, and to arbitrary hybrid dyad $\ketbra{\alpha}{\beta}\otimes\ketbra{\psi}{\phi}$ between a qumode and an arbitrary system. 
			
			\section{Noise sources}
			Bosonic (\textsc{b}) loss, quantum-limited amplification, thermal noise, Gaussian displacement noise~(GDN) of rates $\eta$, and qubit~(\textsc{q}) depolarizing and damping channels of strengths $\eta^\prime$ and $p$ respectively, are given by
			\begin{align}
				\mathcal{N}_\mathrm{loss}[\rhob]
				=&\,\sum^\infty_{l=0}\dfrac{\mu^l}{l!}(1-\mu)^{\frac{a^\dag a}{2}}\,a^l\,\rhob\,a^{\dag\,l}\,(1-\mu)^{\frac{a^\dag a}{2}}\,,\nonumber\\
				\mathcal{N}_\mathrm{amp}[\rhob]
				=&\,\dfrac{1}{G}\sum^\infty_{l=0}\dfrac{(1-G^{-1})^l}{l!}\,a^{\dag l}\,G^{-\frac{a^\dag a}{2}}\,\rhob\,G^{-\frac{a^\dag a}{2}}\,a^{l}\,,\nonumber\\
				\mathcal{N}_\mathrm{therm}[\rhob]=&\,\mathcal{N}_\mathrm{amp}[\mathcal{N}_\mathrm{loss}[\rhob]]\left(\substack{\displaystyle G:=1+\eta\,\bar{n}\\ \displaystyle\mu:=1-\frac{1-\eta}{1+\eta\,\bar{n}}}\right),\nonumber\\
				\mathcal{N}_\mathrm{GDN}[\rhob]=&\,\mathcal{N}_\mathrm{amp}[\mathcal{N}_\mathrm{loss}[\rhob]]\left(\substack{\displaystyle G:=1/\eta\\ \displaystyle\mu:=\eta}\right),\nonumber\\
				\mathcal{N}_\mathrm{dep}[\rhoq]
				=&\,(1-\etad)\,\rhoq+\dfrac{\etad}{3}\sum^3_{j=1}\sigma_j\,\rhoq\,\sigma_j\,\,\nonumber\\
				\mathcal{N}_\mathrm{damp}[\rhoq]=&K_0\rhoq K_0^\dagger+K_1\rhoq K_1^\dagger\,.
				\label{eq:noise}
			\end{align}
			where $\bar{n}$ is the mean excitation and $K_0:=\ketbra{0}{0}+\sqrt{1-p}\ketbra{1}{1}$ with $K_{1}:=\sqrt{p}\ketbra{1}{1}$ and $\sqrt{p}\ketbra{0}{1}$, for the phase and amplitude damping noise respectively. 
			
			\section{Normal and antinormal forms}
			The antinormally ordered forms of functions of the number operator are obtained using,
			\begin{align}
				\vdots F(\AdA)\vdots\ket{m}=&\sum_{n=0}^\infty c_n a^n a^{\dagger n} \ket{m}\nonumber\\
				=&\ket{m}\sum_{n=0}^\infty c_n \underbrace{\frac{(m+n)!}{m!}}
				_{\mathclap{\footnotesize{=(-1)^n\left(\frac{\D}{\D x}\right)^n x^{-(m+1)} \bigg|_{x=1}}}}\nonumber\\
				=&\sum_{n=0}^\infty c_n\left(-\frac{\D}{\D x}\right)^n x^{-(\AdA+1)}\bigg|_{x=1}\ket{m}.\nonumber\\
				\implies\vdots F(\AdA)\vdots=&F\left(-\frac{\D}{\D x}\right)\,x^{-(\AdA+1)}\bigg|_{x=1}.
			\end{align}
			Similar calculations for the normally ordered forms give,
			\begin{align}
				:F(\AdA):\,=&F\left(\frac{\D}{\D x}\right)\,x^{\AdA}\bigg|_{x=1}.
			\end{align}
			These lead to the well-known normally and antinormally forms of the exponentials of the number operators,
			\begin{align}
				\E{\lambda \AdA}=&:\E{\AdA(\E{\lambda}-1)}:\,=\E{-\lambda}\vdots\E{(1-\E{-\lambda})\AdA}\vdots
			\end{align}
			and their rearrangements,
			\begin{align}
				:\E{\lambda\AdA}:\,=(1+\lambda)^{\AdA},\qquad\vdots\E{\lambda\AdA}\vdots=(1-\lambda)^{-\AdA-1}.
			\end{align}
			
			\section{Composite amplitude and phase damping noise}
			
			In this section, we derive the composition of amplitude and phase damping noise of equal noise parameters $p$.
			
			The DV amplitude damping channel is rewritten as
			\begin{align}
				&\mathcal{N}_{\text{damp}}^{\text{(amp)}}[\rhoq]=K_0\rho_{\textsc{q}} K_0^\dagger + K_1^{(\text{amp})}\rho_{\textsc{q}} K_1^{\dagger({\text{amp}})}\nonumber\\
				&=(\ketbra{0}{0}+ \sqrt{1-p}\ketbra{1}{1})\rho_{\textsc{q}}(\ketbra{0}{0}+ \sqrt{1-p}\ketbra{1}{1})+p\ketbra{0}{1}\rho_{\textsc{q}}\ketbra{1}{0}\nonumber\\
				&=\frac{1}{2}\bigg\{
				(1-\frac{p}{2}+\sqrt{1-p})\rhoq+(1-\frac{p}{2}-\sqrt{1-p})\sigma_3\rhoq\sigma_3+\frac{p}{2}\bigg[
				\{\rhoq,\sigma_3\}+\sigma_1\rhoq\sigma_1+\sigma_2\rhoq\sigma_2+\I\bigg(\sigma_2\rhoq\sigma_1-\sigma_1\rhoq\sigma_2\bigg)
				\bigg]
				\bigg\}.
			\end{align}
			
			Similarly, the DV phase damping channel is given by,
			\begin{align}
				&\mathcal{N}_{\text{damp}}^{\text{(ph)}}[\rhoq]=K_0\rho_{\textsc{q}} K_0^\dagger + K_1^{(\text{ph})}\rho_{\textsc{q}} K_1^{\dagger({\text{ph}})}\nonumber\\
				&=(\ketbra{0}{0}+ \sqrt{1-p}\ketbra{1}{1})\rho_{\textsc{q}}(\ketbra{0}{0}+ \sqrt{1-p}\ketbra{1}{1})+p\ketbra{0}{0}\rho_{\textsc{q}}\ketbra{0}{0}\nonumber\\
				&=\left(\frac{1}{2}+\frac{\sqrt{1-p}}{2}\right)\rhoq+\left(\frac{1}{2}-\frac{\sqrt{1-p}}{2}\right)\sigma_3\rhoq\sigma_3.
			\end{align}
			
			The composite DV amplitude and phase damping channel of equal noise strength is then given by,
			\begin{align}
				&\mathcal{N}_{\text{damp}}[\rhoq]:=\mathcal{N}_{\text{damp}}^{\text{(ph)}}\circ\mathcal{N}_{\text{damp}}^{\text{(amp)}}[\rhoq]=\frac{1}{4} \big\{
				c_+\rhoq+c_-\sigma_3\rhoq\sigma_3+p\big[
				\sigma_1\rhoq\sigma_1+\sigma_2\rhoq\sigma_2+\{\rhoq,\sigma_3\}+\I(\sigma_2\rhoq\sigma_1-\sigma_1\rhoq\sigma_2)
				\big]
				\big\}, 
				\label{eq:damp}
			\end{align}
			in a single parameter expression
			where 
			\begin{align}
				c_{\pm}:=&(1+\sqrt{1-p})\left(1-\frac{p}{2}\pm\sqrt{1-p}\right)+(1-\sqrt{1-p})\left(1-\frac{p}{2}\mp\sqrt{1-p}\right).
			\end{align}
			The two damping channels commute, so the ordering is irrelevant.

			\section{Averaging over random states}
			
			The moments of pure Haar random states are given by~\cite{Harrow2013:Church, Mele2024:Introduction}
			\begin{align}
				\AVG{U\sim \mu_{\textsc{h}}}{U^{\otimes t}\ketbra{\text{\o}^{\otimes t}}{{\text{\o}}^{\otimes t}}U^{\dagger \otimes t}}=\frac{1}{{{d+t-1}\choose{t}}} P_{\mathrm{sym}}=\frac{t!}{d^{\bar{t}}}P_{\mathrm{sym}},
			\end{align}
			where $\ket{{\text{\o}}}$ is an irrelevant pure fiducial symmetric state, where $P_{\mathrm{sym}}=\frac{1}{t!}\sum_{\pi \in S_t}P_d(\pi)$ is the projector onto the space of symmetric permutations $\pi$ of $t$ systems of dimension $d$, $P_d(\pi)$ is the corresponding representation of the permutation on $(\mathbb{C}^d)^{\otimes t}$ and $d^{\bar{t}}=d(d+1)\cdots(d+t-1)$ are the rising factorials.
			
			Therefore, 
			\begin{align}
				&\AVG{U\sim\mu_{\textsc{h}}}{\tr{U^{ \otimes t}\ketbra{\text{\o}^{\otimes t}}{\text{\o}^{\otimes t}}U^{\dagger \otimes t}(M_1\otimes M_2\otimes \cdots \otimes M_t)}}\nonumber\\
				&=\frac{1}{d^{\bar{t}}}\sum_{\pi \in S_t}
				\tr{P_d(\pi)^\dagger(M_1\otimes M_2\otimes \cdots \otimes M_t)}\nonumber\\
				&=\frac{1}{d^{\bar{t}}} \sum_{\pi \in S_t}\prod_{c \in \operatorname{cycles}(\pi)}\mathop{\mathrm{tr}}\big\{\prod_{j \in c} M_j\big\}
			\end{align}
			
			We assume the computational logical states to be orthogonal in the bosonic codespace defined by the normalized identity $\SL:=\big(\ketbra{0_\textsc{l}}{0_\textsc{l}}+\ketbra{1_\textsc{l}}{1_\textsc{l}}\big)/2$ and that
			\begin{align}
				M_j:=&\begin{pmatrix}
					\opinner{0_\textsc{L}}{X_j}{0_\textsc{L}} & \opinner{0_\textsc{L}}{X_j}{1_\textsc{L}}\\
					\opinner{1_\textsc{L}}{X_j}{0_\textsc{L}} & \opinner{1_\textsc{L}}{X_j}{1_\textsc{L}}
				\end{pmatrix}
			\end{align}
			with $X_j$ being the bosonic operators. 
			
			Therefore, applying the previous results to three cases, $t = 1,2,3$, and defining the logical $\rho_{\textsc{l}}:=U^\dagger\ket{{\text{\o}}}\!\!\bra{{\text{\o}}}U$, we get
			
			\begin{align}
				\AVG{\rho_L}{\Tr{\rho_{\textsc{l}} M_1}}&=\frac{1}{2}\Tr{M_1},\nonumber\\
				\AVG{\rho_{\textsc{l}}}{\Tr{\rho_{\textsc{l}} M_1 \rho_{\textsc{l}} M_2}}&=\AVG{\rho_{\textsc{l}}}{\Tr{\rho_{\textsc{l}}M_1}\Tr{\rho_{\textsc{l}}M_2}}=\frac{1}{6}\left(\Tr{M_1M_2}+\Tr{M_1}\Tr{M_2}\right),\nonumber\\
				\AVG{\rho_{\textsc{l}}}{\Tr{\rho_{\textsc{l}}M_1\rho_{\textsc{l}}M_2}\Tr{\rho_{\textsc{l}}M_2}}&=\AVG{\rho_{\textsc{l}}}{\Tr{\rho_{\textsc{l}}M_1\rho_{\textsc{l}}M_2\rho_{\textsc{l}}M_3}}\nonumber\\
				&=\frac{1}{24}\big(\Tr{M_1}\Tr{M_2}\Tr{M_3}+\Tr{M_1M_2M_3}+\Tr{M_3M_2M_1}\nonumber\\
				&\indent\indent+\Tr{M_1}\Tr{M_2M_3}+\Tr{M_2}\Tr{M_3M_1}+\Tr{M_3}\Tr{M_1M_2}\big)\nonumber\\
				&=\frac{1}{12}\big(\Tr{M_1}\Tr{M_2}\Tr{M_3}-\Tr{M_3[M_1,M_2]}+2\Tr{M_1M_2M_3}\big)
			\end{align}

			With the identities and definitions set up, we have
			$\Tr{M_1}=2\tr{X_1\SL}$, $\Tr{M_1M_2}=4\tr{X_1\SL M_2 \SL}$, and $\Tr{M_1M_2M_3}=8\tr{X_1\SL X_2\SL X_3\SL}$
			and therefore,
			\begin{align}
				\AVG{\rho_{\textsc{l}}}{\Tr{\rho_{\textsc{l}} M_1}}&=\tr{M_1\SL}\nonumber\\
				\AVG{\rho_{\textsc{l}}}{\Tr{\rho_{\textsc{l}} M_1\rho_{\textsc{l}} M_2}}&=\frac{2}{3}\argp{\tr{X_1\SL X_2\SL}+\tr{X_1\SL}\tr{X_2\SL}}\nonumber\\
				\AVG{\rho_{\textsc{l}}}{\Tr{\rho_{\textsc{l}} X_1\rho_{\textsc{l}} X_2\rho_{\textsc{l}} C}}
				&=\frac{2}{3}\bigg(\tr{X_1\SL} \tr{X_2\SL}\tr{X_3\SL}-\tr{X_3\SL[X_1\SL,X_2\SL]}
				+2\tr{X_1\SL X_2\SL X_3\SL}\bigg).
			\end{align}
			
			\section{Performance of photon loss and thermal noise suppression under depolarizing ancilla noise}
			\begin{figure}[H]
				\centering
				\includegraphics[width=0.75\columnwidth]{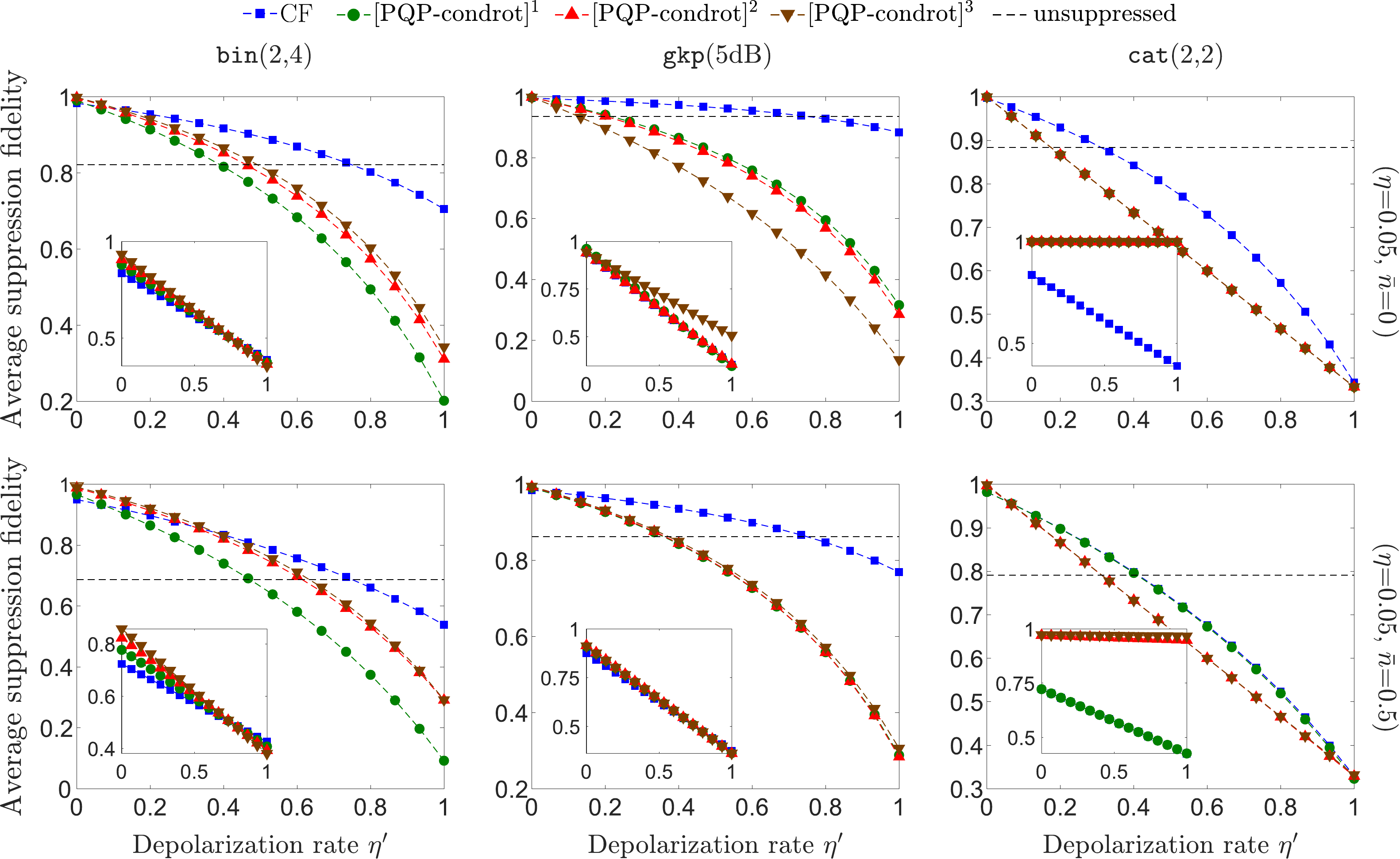}
				\caption{\label{fig:PQPcondrot_dep}The performance of the conditional-Fourier interferometer is more resilient to the DV depolarizing noise as well. Its average suppression performance remains somewhat impervious to uncalibrated ancilla depolarizing noise~Eq.~\eqref{eq:noise}, contrary to a series of conditional displacement gates and conditional rotation gates~([PQP-condrot]$^L$) [see Fig.~1~(b) of the main text], numerically optimized for known noise parameters for photon loss~($\eta=0.05$) and thermal noise~($\eta=0.05$, $\bar{n}=0.5$). Insets show average success~probability.}
			\end{figure}
			
			\section{Series of Jaynes--Cummings interactions and conditional rotations}
			\begin{figure}[H]
				\centering
				\includegraphics[width=0.75\columnwidth]{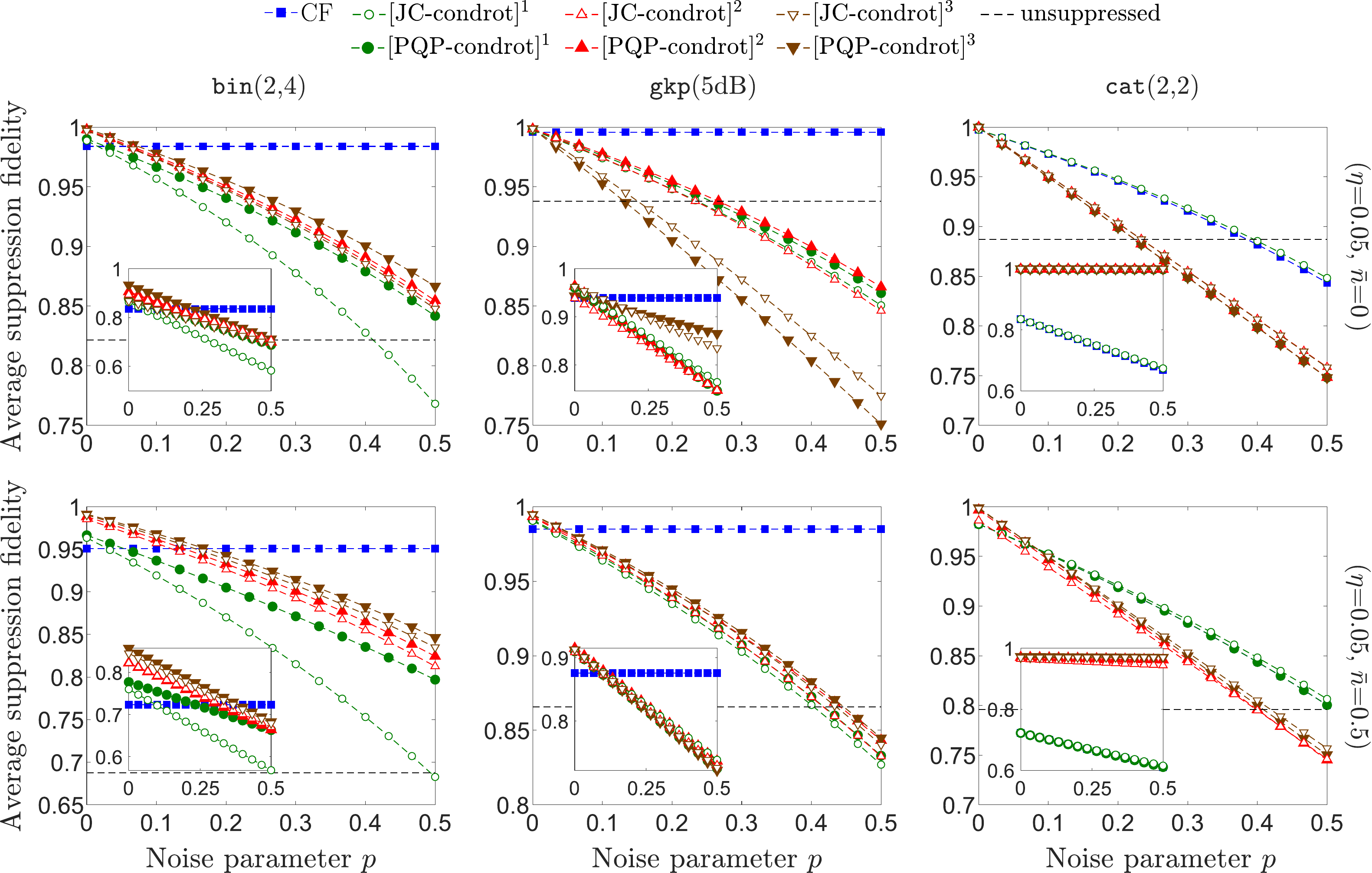}
				\caption{\label{fig:JCcondrot_dep} A series of conditional displacement gates and conditional rotation gates~([PQP-condrot]$^L$) [see Fig.~1~(b) of the main text] compared to an alternate, series of Jaynes--Cummings interactions and conditional rotation gates~([JC-condrot]$^L$), both numerically optimized for known noise parameters for photon loss~($\eta=0.05$) and thermal noise~($\eta=0.05$, $\bar{n}=0.5$). Insets show average success~probability.}
			\end{figure}

	\end{document}